\documentclass[reprint,aps,prx,showpacs,amsmath,amssymb,superscriptaddress,showkeys,longbibliography]{revtex4-2}

\usepackage{bm}
\usepackage{float}
\usepackage{mathtools}
\usepackage{multirow}
\usepackage{bbold}
\usepackage{bbm}
\usepackage{braket}
\usepackage{bbold}
\usepackage{amsthm}
\usepackage{mathrsfs}
\usepackage[usenames]{xcolor}
\usepackage{gensymb}
\usepackage{appendix}
\usepackage{graphicx}
\usepackage[caption=false]{subfig}
\usepackage{tikz}
\usepackage{dcolumn}

\usepackage{color}

\usepackage{soul} 
\usepackage{hyperref}

\newcommand{\METHOD}{\textrm{WTG+FET }}
\newcommand{\SU}{\textrm{SU }}

\begin{document}
\def\mean#1{\left< #1 \right>}
\title{Dynamics of two-dimensional open quantum lattice models with tensor networks}

\author{C. Mc Keever}
\affiliation{Department of Physics and Astronomy, University College London,
Gower Street, London, WC1E 6BT, United Kingdom}
\author{M. H. Szyma\'{n}ska}
\affiliation{Department of Physics and Astronomy, University College London,
Gower Street, London, WC1E 6BT, United Kingdom}

\date{\today}

\begin{abstract}

Being able to describe accurately the dynamics and steady-states of driven and/or dissipative but quantum correlated lattice models is of fundamental importance in many areas of science: from quantum information to biology. An efficient numerical simulation of large open systems in two spatial dimensions is a challenge.
In this work, we develop a tensor network method, based on an infinite Projected Entangled Pair Operator (iPEPO)  ansatz, applicable directly in the thermodynamic limit. We incorporate techniques of finding optimal truncations of enlarged network bonds by optimising an objective function appropriate for open systems. 
Comparisons with numerically exact calculations, both for the dynamics and the steady-state, demonstrate the power of the method. In particular, we consider dissipative transverse quantum Ising and driven-dissipative hard core boson models in non-mean field limits, proving able to capture substantial entanglement in the presence of dissipation. 
Our method enables to study regimes which are accessible to current experiments but lie well beyond the applicability of existing techniques.
\end{abstract}


\maketitle



\section{Introduction}

In recent experiments across a variety of architectures, the ability to sustain quantum correlations in a dissipative environment and study the evolution of strongly interacting many-body lattice systems in a precisely controlled manner, has progressed enormously. Among these experimental platforms are cavity \cite{walther2006cavity,reiserer2015cavity} and circuit \cite{schmidt2013circuit,houck2012chip,kollar2019hyperbolic,carusotto2020photonic} QED systems, arrays of coupled optical cavities \cite{carusotto2009fermionized,umucalilar2012fractional,grujic2012non} or of quantum dots \cite{kasprzak2010up}, hybrid systems \cite{jin2015proposal}, polariton lattices \cite{amo2016exciton,schneider2016exciton,kim2011dynamical,tanese2013polariton,baboux2016bosonic,klembt2017polariton,whittaker2018exciton,dusel2020room} and certain implementations of ultracold atoms \cite{brennecke2007cavity}.

\begin{figure}[b!]
  	\includegraphics[width=\linewidth]{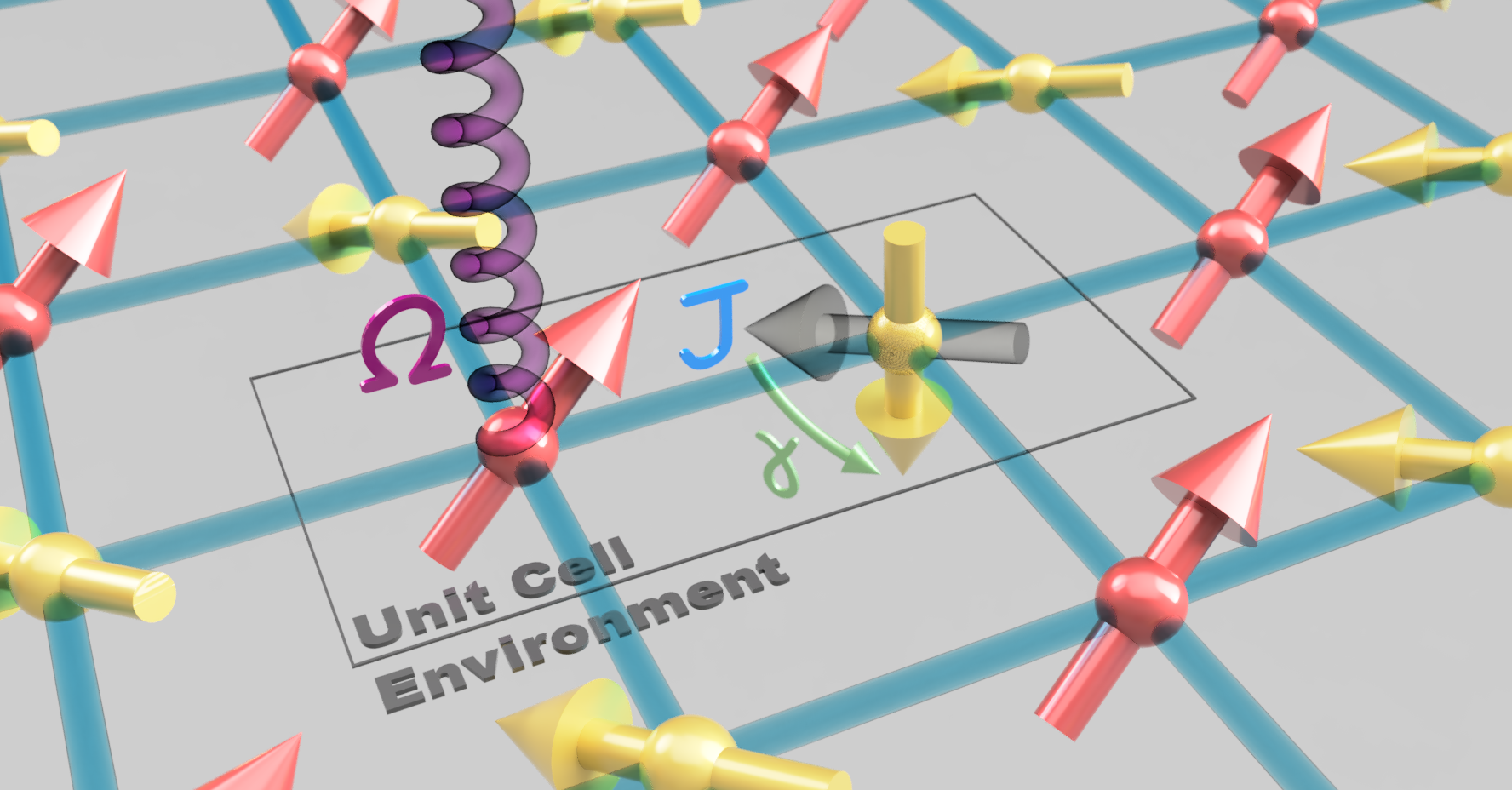}
  	\caption{\label{fig:main} \textbf{An open quantum lattice model of interacting spins.} Nearest neighbour spins are coupled via a hopping  $J$ and interact with an external \textit{bath} via a coherent drive $\Omega$ and/or a dissipative process $\gamma$. The open system can be modelled by describing the \textit{unit cell} and its \textit{environment} using a tensor network. }
\end{figure}

In the modelling of these systems, the inclusion of degrees of freedom which are external to the lattice, such as a driving field or a bath of oscillators, requires extending the description from a \textit{closed} to an \textit{open} quantum lattice model, as illustrated in Fig. \ref{fig:main}. Open quantum systems are often well described by a Lindblad master equation \cite{breuer2002theory} which facilitates the study of a range of collective phenomena including non-equilibrium criticality \cite{sieberer2013dynamical,lee2012collective,jin2016cluster,nissen2012nonequilibrium,marino2016driven,fitzpatrick2017observation}, quantum chaos \cite{gao2015observation,fernandez2014lattice} and time-crystallinity \cite{iemini2018boundary,tucker2018shattered,zhu2019dicke}, many of which have no counterparts in closed systems at equilibrium. However, to better understand, control and utilise the dissipative non-equilibrium dynamics of correlated quantum systems, simulation techniques which are scalable to large lattices are still missing, especially in higher dimensions.

The investigation of large many-body quantum systems is hindered by the exponential growth of the Hilbert space. As the size of the system increases, solving the Lindblad master equation exactly using methods such as diagonalization of the Liouvillian or averaging over ensembles of exact quantum trajectories \cite{plenio1998quantum,dalibard1992wave,tian1992quantum} quickly become infeasible. To simplify the problem, many have resorted to a mean field type approximation \cite{nissen2012nonequilibrium,jin2013photon,jin2014steady,lee2013unconventional,le2013steady,le2014bose,tomadin2010signatures,diehl2010dynamical} in which correlations between small individual subsystems are approximated by an average field. This simplification, however, may often give qualitatively incorrect results in regions where inter-subsystem correlations become important --- for instance, near criticality. Moreover, key aspects such as entanglement and quantum information cannot be treated at this level.
Progressing beyond mean field approximations should therefore involve the systematic inclusion of correlations between subsystems in a controlled and tractable manner.

In this vein, phase space methods such as those based on the Wigner \cite{wigner1997quantum}, Positive-P \cite{Drummond80} and Q \cite{cahill1969density} representations attempt to find classical stochastic processes for which the hierarchy of couple moments is a good approximation to that of the quantum problem. For highly non-linear problems, phase space techniques often fail dramatically in important regimes \cite{gilchrist1997positive,deuar2020scalable,schachenmayer2015many}. Cluster based methods \cite{jin2016cluster,biella2018linked} separate large lattices into small clusters and capture correlations within lattice sites belonging to each cluster, an approach which can become inaccurate when correlation lengths exceed cluster sizes. Variational approaches \cite{weimer2015variational1,weimer2015variational2} based on the parametrization of the state in terms of a suitable functional and their optimisation relies on good intuition, which may not be available for some problems. Recently, methods based on neural networks and the variational minimization of an appropriate cost function \cite{hartmann2019neural,vicentini2019variational,yoshioka2019constructing} have provided an interesting proof of concept, however, like most of these methods, they are restricted to small system sizes or may fail to capture long range correlations.

A different approach is to restrict the growth of the system's Hilbert space by retaining only the most important correlations or most probable states \cite{finazzi2015corner}. Tensor network (TN) methods \cite{orus2014practical} belong to this class. Here, truncation of Hilbert space is controlled by the so called bond dimension (usually denoted $D$ or $\chi$) of indices which connect a set of tensors representing the quantum state. In the context of closed quantum many-body systems, the significant success of TN methods is underpinned by an area law in the growth of entanglement entropy possessed by ground states of gapped Hamiltonians \cite{hastings2007area}. For open systems the picture is much less clear. In particular it is not obvious whether transient or steady states can be efficiently represented by a TN. Nevertheless, in the context of dissipative or driven-dissipative systems, we can reasonably expect that in many cases, dissipative processes should curtail the growth of entanglement and limit correlations generated by entangling dynamics.

Despite this expectation, TN algorithms for open systems \cite{werner2016positive,cui2015variational,mascarenhas2015matrix,verstraete2004matrix,gangat2017steady} have mostly been restricted to one-dimensional lattices where the simple geometry plays a central role in the algorithm. In dimensions greater than one, progress has been limited. The work of \cite{kshetrimayum2017simple} introduced the Infinite Projected Entangled Pair Operator (iPEPO) to represent the mixed state of an infinite periodic two-dimensional square lattice and employed the so called \textit{simple update} (SU) algorithm to apply Lindblad dynamical map evolving the system in real time towards a steady state. Although SU is efficient, in order to integrate the equation of motion it isolates a subsystem --- for example one unit cell --- from the rest of the lattice and applies the dynamical map to the subsystem in isolation until a steady state is reached. It has been questioned whether this approach can produce accurate results and there are concerns over the convergence of this method in non-mean-field regimes \cite{kilda2020ipepo}. While algorithms going beyond SU exist for closed and finite temperature systems \cite{czarnik2019time,czarnik2015projected,phien2015infinite}, advancing beyond the SU approach in the driven-dissipative context remains undeveloped.

In this paper we devise a \emph{new} TN method to accurately simulate time dynamics and steady states of many-body quantum lattice models in \emph{two spatial dimensions} and directly in the \emph{thermodynamic limit}. The method uses the iPEPO as an ansatz for the mixed state of the open system and incorporates techniques inspired by those presented in \cite{evenbly2018gauge} --- Full Environment Truncation (FET) and fixing the network to Weighted Trace Gauge (WTG) --- to calculate accurate time dynamics and steady state solutions of open quantum lattice models. The central step in the algorithm involves finding an optimal truncation of enlarged bonds with respect to an objective function appropriate for mixed quantum states. 

The method successfully reproduces numerically exact calculations for both dynamics and steady-states while also agreeing with results obtained using the so called Corner Space Renormalization method of \cite{finazzi2015corner}. Importantly, it performs well in
non-mean field limits, proving able to capture substantial correlations in the presence of dissipation and therefore enabling the study of regimes which are accessible to current experiments but lie well beyond the applicability of existing techniques.

The paper is organised as follows. In section \ref{section:the_algorithm} we describe the algorithm including a brief introduction to the Lindblad master equation and the TN ansatz. As a benchmark we calculate time dynamics of a dissipative transverse quantum Ising model in section \ref{subsection:IsingModel} and find that the systematic inclusion of correlations - controlled by the TN bond dimension - coupled with the incorporation of the unit cell's environment when truncating enlarged bonds yields results which agree very well with the exact dynamics. Furthermore we demonstrate the applicability of the algorithm outside the exactly solvable regime. In section \ref{subsection:comparison_with_simple_update} we show that the FET method outperforms the SU method by finding more optimal truncations of enlarged bonds by removing redundant internal correlations in the network. Finally in section \ref{subsection:drive_dissipative_XY_model} we show that lattice models with drive and dissipation can also be treated using this method and compare steady state results for a driven-dissipative hard core boson model with literature values. In section \ref{section:perspectives} we conclude with a short discussion.

\section{The Algorithm}\label{section:the_algorithm}

\subsection{Master Equation}\label{subsection:master_equation}
The goal of the algorithm is to calculate time dynamics and steady states of translationally invariant two-dimensional quantum lattice models, which interact with a bath via a Lindblad master equation (\ref{eqn:lindblad}) $(\hbar=0)$
\begin{equation}
\label{eqn:lindblad}
\frac{d \hat\rho}{dt} = \mathcal{\hat L}\left( \hat\rho \right) = -i\left[ \mathcal{\hat H}, \hat\rho \right] + \mathcal{\hat D}\left( \hat\rho \right),
\end{equation}
where $\mathcal{\hat{H}}$  governs the coherent dynamics of the \textit{system} and the dissipator $\mathcal{\hat D}$, which models the coupling of the system to its \textit{bath} has the form 
\begin{equation}
\label{eqn:dissipator}
\mathcal{\hat D}(\hat \rho) = \sum_{\alpha} \left( \hat L_{\alpha} \hat\rho \hat L_{\alpha}^{\dagger}  - \frac{1}{2} \{ \hat L_{\alpha}^{\dagger} \hat L_{\alpha},\hat \rho \}\right),
\end{equation}
with  $\hat L_{\alpha}$ being the Lindblad operators. We focus on the case of time-independent nearest neighbour Hamiltonians such that $H$ can be decomposed as a sum of Hermitian operators which act non-trivially on at most two nearest neighbour lattice sites.  Although the algorithm allows for up to two-local dissipators, for simplicity, we focus only on local coupling to the environment such that each Lindblad operator acts on one site only and respects the translational invariance of the Hamiltonian.

\subsection{TN Ansatz}\label{subsection:tensor_network_ansatz}
\begin{figure}[!ht]
 \includegraphics[width=\linewidth]{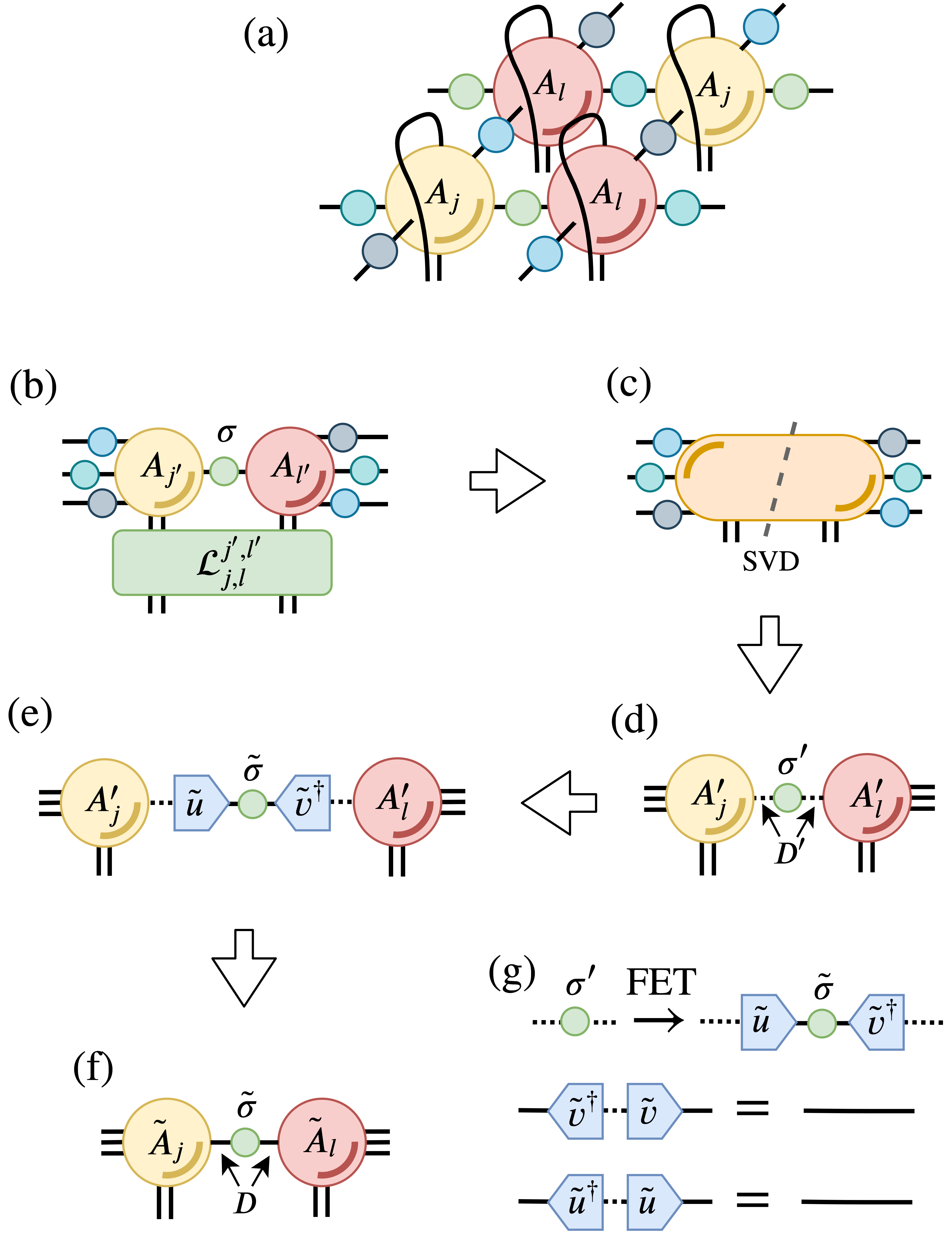}
  \caption{\textbf{Main steps in the time evolution algorithm.} (a) The vectorized form of the iPEPO. (b) The contraction of $A_{j'}$ and $A_{l'}$ with the dynamical map $\mathcal{L}$. (c) Singular value decomposition (SVD) of $e^{\tau \mathcal{L}}A_jA_l$. (d) The $D'$ singular values of relative tolerance greater than $\epsilon_{D'}$ are retained in the diagonal bond matrix $\sigma'$. (e) The isometries $\tilde u$ and $\tilde v$ truncate the enlarged bond from $D'$ to $D$ giving the new bond matrix $D$. (f) The updated tensors $\tilde A_j$ and $\tilde A_l$. (g) The Full Environment Truncation (FET) algorithm is used to find the isometries $\tilde u$ and $\tilde v$ which maximize the fidelity between truncated and untruncated bonds.}
  \label{fig:basic_tensors}
\end{figure}

We represent the system's density matrix $\rho\left(t\right)$ as an infinite Projected Entangled Pair Operator (iPEPO). The iPEPO is composed of a network of tensors $\{A_j\}$, where we associate each node $j$ of the network with one site of the square lattice shown in Fig. \ref{fig:basic_tensors} (a). To reflect the translational invariance of the system and to simplify the algorithm, we use a pair of independent tensors $A_j$ and $A_l$ to represent the unit cell. The infinite system is the repetition of this unit cell over the two-dimensional plane. Each sixth-rank tensor $A$ has a pair of physical indices of dimensions $d$ and a set of four bond indices of dimension $D$, reflecting the coordination number $z=4$ associated with a square lattice. The physical dimension $d$ corresponds to the dimension of the local Hilbert space at each lattice site ($d=2$ for the two-level spin), whereas $D$ is a variational parameter which controls the accuracy of the ansatz. It is convenient to use the vectorized form of the density operator, which at the level of the iPEPO corresponds to vectorization of the pair of local Hilbert space indices as shown in Fig. \ref{fig:basic_tensors} (a) and has the effect of transforming the iPEPO into the form of a infinite Projected Entangled Pair State (iPEPS) commonly used in TN algorithms for two-dimensional closed systems \cite{orus2014practical}. Finally, to each unique bond we associate a bond matrix $\sigma$.

As with other algorithms based on Matrix Product Operators (MPOs), the PEPO ansatz is not inherently positive and therefore not all PEPOs represent physical states. For the present case of an \textit{infinite} PEPO (iPEPO) we do not have access to the full spectrum of eigenvalues and it has been shown for the case of MPOs that the problem of deciding whether a given iMPO represents a physical state in the thermodynamic limit is provably undecidable \cite{kliesch2014matrix}. We therefore rely on the positivity of the dynamical map to maintain the physicality of the iPEPO throughout the time evolution and find in practice that, in most cases, the reduced density matrices calculated from the iPEPO are physical. 

We refer to all of the spins in the system which are not part of the \textit{unit cell} as its \textit{environment} (see Fig. \ref{fig:main}.), not to be confused with system's \textit{bath} which is accounted for in the Lindblad master equation (\ref{eqn:lindblad}). Since the system is infinite, we represent the environment approximately by associating to each tensor in the unit cell an \textit{effective environment} $\mathcal{E}$. $\mathcal{E}$ is itself made up of a set of tensors including four corner transfer matrices $C_{\mu \nu}$ and four half row or half column tensors $T_{\mu}$, where the labels $\mu$ and $\nu$ take the appropriate first letter of \textbf{l}eft, \textbf{r}ight, \textbf{u}p and \textbf{d}own as illustrated in Fig. \ref{fig:bond_environment}. (a-b). 

We consider two distinct types of effective environment. The ``trace effective environment'' $\mathcal{E}^{tr}$ of Fig. \ref{fig:bond_environment} (a) is calculated by first tracing over the local Hilbert space dimensions $d$ of the tensors at each node of the network giving the set of fourth-rank tensors $\{ a^{tr}_j \}$ as shown in Fig. \ref{fig:CTMRG} (a). We use $\mathcal{E}^{tr}$ to calculate the reduced density matrices of the system. Secondly, the ``Hilbert-Schmidt effective environment''  $\mathcal{E}^{hs}$ (Fig. \ref{fig:bond_environment} (a)) is that formed by first contracting $a^{hs}_j = \vec{A}_j\vec{A}^{\dagger}_j$ giving the Hilbert-Schmidt inner product of the tensor $\vec{A}_i$ with itself, where all bond indices $\{D\}$ are left open as shown in Fig. \ref{fig:CTMRG} (c). $\mathcal{E}^{hs}$ is used during the algorithm to calculate an optimal truncation of enlarged bond dimensions as discussed in section \ref{subsection:truncation_of_enlarged_bonds}. In both cases we calculate the effective environment using a corner transfer matrix method \cite{baxter1968dimers,baxter1978variational,nishino1996corner,nishino1997corner,orus2009simulation,corboz2010simulation}. In particular, we use a variant of the Corner Transfer Matrix Renormalization Group (CTMRG) algorithm \cite{fishman2018faster} which makes use of an intermediate SVD to improve stability, details of which are given in Appendix \ref{appendix:calculating_the_environment}.

\subsection{Time Evolution}\label{section:time_evolution}
To calculate dynamics and find a TN representation of the steady state we use a time evolving block decimation (TEBD) algorithm. The time evolution is obtained by application of the dynamical map $\rho_t = e^{t \mathcal{L}}\rho_0$. In principle it may also be possible to find the steady state directly by searching for the ground state of the Hermitian operator $\rightarrow\mathcal{L}^{\dagger}\mathcal{L}$, for example, via imaginary time evolution. However, in general, $\mathcal{L}^{\dagger}\mathcal{L}$ is a highly non-local operator and is therefore not straightforward to implement using standard techniques for an infinite systems \footnote{An extension of the hybrid method presented in \cite{gangat2017steady} to two-dimensions appears straightforward, however increasing the local Hilbert space dimension of the iPEPO to accommodate for example an 8-local nearest neighbour operator would introduce significant computational cost.}. Finally, access to the transient dynamics is often of direct interest in many physical contexts.

The dynamical map $e^{t \mathcal{L}}$ is approximated by a set of Trotter layers as it is common in algorithms based on TEBD. In particular, consider the evolution of the state from a time $t$ to a short time later $t+\tau$, then, in vectorized notation, where we note that the density matrix is vectorized column-by-column, the dynamical map takes the form
\begin{equation}
\rho(t+\tau) = e^{\tau \mathcal{L}}\rho(t).
\end{equation}
The Liouvillian superoperator $\mathcal{L}$ is two-local and can therefore be written as a sum of superoperators acting on nearest neighbours of the square lattice, where the labels $\alpha$ and $\beta$ correspond to the coordinates of the lattice site $j$ and $l$ respectively. The full Liouvillian takes the form 
\begin{equation}
\mathcal{L} = \sum_{\langle\alpha,\beta\rangle}\mathcal{L}_{\alpha,\beta}=\sum_{\langle\alpha,\beta\rangle}\mathcal{H}_{\alpha,\beta} + \mathcal{D}_{\alpha,\beta}.
\end{equation}
The Hamiltonian part of the evolution is included in the superoperator $\mathcal{H}$ and the dissipative part in the superoperator $\mathcal{D}$ each are constructed as shown in equations (\ref{eqn:ham_vec}) and (\ref{eqn:luv_vec}) respectively: 
\begin{equation}\label{eqn:ham_vec}
\mathcal{H}_{\alpha,\beta} = -i \left(\mathbb{I}_{\alpha,\beta}\otimes H_{\alpha,\beta} - H^{T}_{\alpha,\beta}\otimes \mathbb{I}_{\alpha,\beta} \right),
\end{equation}
\begin{equation}\label{eqn:luv_vec}
\mathcal{D}_{\alpha,\beta} = \frac{1}{2}(2 L_{\alpha,\beta}^* \otimes L_{\alpha,\beta} - I_{\alpha,\beta} \otimes L^{\dagger}L_{\alpha,\beta} - L^T L_{\alpha,\beta}^* \otimes I_{\alpha,\beta}).
\end{equation}
We then split the vectorized operators in the exponent into those acting on even and odd pairs of lattice sites along both the $x$ and $y$ lattice dimensions, giving four sets of vectorized operators $\mathcal{L}_x^e$, $\mathcal{L}_x^o$, $\mathcal{L}_y^e$ and $ \mathcal{L}_y^o$ where 
\begin{equation}
\mathcal{L}_r^e=\sum \mathcal{L}_{2r,2r+1}, \quad \mathcal{L}_r^o=\sum \mathcal{L}_{2r-1,2r},
\end{equation}
which allows us to decompose $e^{t\mathcal{L}}$ into a set of layers via a \textit{Trotter decomposition} with $\tau=t/n$ where $n\gg 1$ is the \textit{Trotter number} with
\begin{equation}
\label{eqn:layers} 
e^{\tau \mathcal{L}} = e^{\tau\mathcal{L}_x^e}e^{\tau\mathcal{L}_x^o}e^{\tau\mathcal{L}_y^e}e^{\tau\mathcal{L}_y^o} + \mathcal{O}(\tau^2).
\end{equation}
Each dynamical map in the decomposition is applied to pairs of nearest neighbour tensors $A_j$ and $A_l$ in turn. We first construct the linear map $\mathcal{L}\left(A_jA_l\right)$ where the linear operator $\mathcal{L}^{j',l'}_{j,l}$ acts on the pair of tensors $A_j$ and $A_l$ such that $A_jA_l$ behaves as a vector in the linear map as illustrated in Fig. \ref{fig:basic_tensors}. (b). By repeated application of this map, an approximation to the tensor $e^{\tau \mathcal{L}}A_jA_l$  (Fig. \ref{fig:basic_tensors} (c)) is calculated using Krylov subspace methods, eliminating the need for explicit calculation of $e^{\tau \mathcal{L}}$, where $\tau$ is a real number for the case of real time evolution. 

To complete the update, the resulting tensor $A'_{j,l}=e^{\tau\mathcal{L}}A_{j}A_{l}$ needs to be decomposed into a new pair of tensors $A'_j$ and $A'_l$, illustrated in Fig. \ref{fig:basic_tensors} (c-d). Typically this is done via singular value decomposition (SVD), where in general, the new bond dimension $D'$ --- equal to the number of singular values associated with the SVD --- will be enlarged $(D'>D)$ and therefore needs to be truncated in an appropriate way for the algorithm to remain efficient, in particular, we would like to truncate $D'$ back to $D$ after each dynamical map.

\subsection{Truncation of Enlarged Bonds}\label{subsection:truncation_of_enlarged_bonds}
For TNs without closed loops (\textit{acyclic}), finding an optimal truncation benefits greatly from the ability to efficiently apply a gauge transformation and re-cast a network to a so called \textit{canonical form}, for details we refer the reader to \cite{orus2014practical}. For TNs with closed loops (\textit{cyclic}) however, such a canonical form cannot be defined uniquely and truncating the enlarged bond in an optimal way is much less straightforward. Moreover, cyclic TNs can host so called \textit{internal} correlations which have no influence on properties of the quantum state but can cause computational problems if they are allowed to accumulate \cite{evenbly2018gauge}.


After applying the dynamical map we choose to decompose the tensors using SVD and truncate the bond irrespective of the state of the environment, leaving a new dimension $D' \geq D$ chosen such that only those singular values greater than some small tolerance $\epsilon_{D'}\ll1$ are retained. We are then left with a bond matrix $\sigma$ with the remaining $D'$ singular values along its diagonal and the tensors $A_i$ and $A_j$ as shown in Fig. \ref{fig:basic_tensors} (d). The final step in the truncation involves replacing $\sigma$ with the product $\tilde{u}\tilde{\sigma}\tilde{v}^{\dagger}$ where $\tilde{u}$ and $\tilde{v}$ are isometries of dimension $(D',D)$ such that $\tilde{u}\tilde{u}^{\dagger}=\tilde{v}\tilde{v}^{\dagger}=I$ and $\tilde{\sigma}$ is a new $D$ dimensional diagonal bond matrix. The enlarged bond is then truncated by contracting $A_j$ and $A_l$ with $\tilde{u}$ and $\tilde{v}$ as illustrated in Fig. \ref{fig:basic_tensors} (e). 
 
To calculate the set $\tilde{u}$, $\tilde{\sigma}$ and $\tilde{v}$ we adapt the Full Environment Truncation (FET) algorithm of \cite{evenbly2018gauge}, which prescribes a method to find the truncation of an internal index of an arbitrary network for closed systems, optimal with respect to a fidelity measure for pure states. In our case, since we are dealing with an open system, we optimize the truncation with respect to an objective function suitable for mixed states. More precisely, we maximize a mixed state fidelity measure between the state $\rho$ in which the enlarged bond dimension is left untruncated and the state $\phi$ in which the same bond has been truncated by $\tilde{u}$, $\tilde{\sigma}$ and $\tilde{v}$. Supposing that a global maximum is found, this procedure finds the isometries which leave $\phi$ as close as possible to $\rho$ with respect to the chosen fidelity measure.

We choose to maximize the fidelity $\mathcal{F}\left(\rho,\phi\right)$, which has the Hilbert-Schmidt inner product of $\rho$ and $\phi$ in its numerator and the geometric mean of their purities $tr(\rho^2)$ and $tr(\phi^2)$ in its denominator \cite{wang2008alternative}
\begin{equation}\label{eqn:fidelity}
\mathcal{F}(\rho,\phi) = \frac{tr(\rho\phi)}{\sqrt{tr(\rho^2)tr(\phi^2)}}.
\end{equation}
Since squaring $\mathcal{F}$ is convex, the $\rho$ and $\phi$ which maximize $\mathcal{F}^2\left(\rho,\phi\right)$ also maximize $\mathcal{F}\left(\rho,\phi\right)$. We therefore construct $\mathcal{F}^2\left(\rho,\phi\right)\textrm{tr}\left(\rho^2\right)$ as a Rayleigh quotient of tensors which can be maximized to find an optimal $\tilde{u}$, $\tilde{\sigma}$ and $\tilde{v}$. Details of the optimization procedure are given in Appendix \ref{appendix:full_environment_truncation}.

Finally, there exists a gauge freedom across the newly truncated bond which we fix to so called Weighted Trace Gauge (WTG) as described in \cite{evenbly2018gauge}. This allows for the recycling of the environment $\mathcal{E}^{hs}$ calculated for use at each FET step of the algorithm as an initial guess for the renormalization procedure (CTMRG in our case) which precedes the following FET step thereby reducing the number of renormalization iterations required at each step. We refer to the algorithm outlined in this section as Full Environment Truncation in Weighted Trace Gauge (\METHOD).

It is straightforward to recover a Simple Update (SU) method by bypassing the FET and WTG steps above and instead choosing both $\tilde{u}\rightarrow\tilde{u}_{su}$ and $\tilde{v}\rightarrow\tilde{v}_{su}$ as $D'\times D$ matrices with all diagonal entries equal to one and all other entries equal to zero and by retaining the $D$ largest singular values of $\sigma'$ in the truncated $\tilde\sigma_{su}$. In general, the set of $\tilde{u}$, $\tilde{v}$ and $\tilde{\sigma}$ we find using FET are not equivalent to $\tilde{u}_{su}$, $\tilde{v}_{su}$ and $\tilde{\sigma}_{su}$ showing that, in the general case, SU does not yield a truncation which is optimal with respect to the objective function we use. A comparison between \SU and \METHOD is made in section \ref{subsection:comparison_with_simple_update}.

\section{Results}\label{section:resutls}

\subsection{Dissipative Transverse Ising Model}\label{subsection:IsingModel}
As a first benchmark of the algorithm we simulate dynamics of a dissipative transverse quantum Ising model with Hamiltonian
\begin{equation}
\label{eqn:DissipativeTransverseIsing}
\hat H = \frac{V}{z} \sum_{\langle j,l\rangle}\hat\sigma^z_j\hat\sigma^z_l  +  \sum_j \frac{h_x}{2}\hat\sigma^x_j,
\end{equation}
where $V$ is the hopping coupling, $h_x$ is the strength of a transverse field and $z$ is the lattice coordination number which we set to $z=4$ for the square lattice. The spins undergo dissipation at a rate $\gamma$ described by local Lindblad jump operators $\hat L_j = \sqrt{\gamma} \frac{1}{2} \left(\hat\sigma^y_j - i\hat\sigma^z_j \right) $, which are the same at each lattice site. For zero transverse field $h^x/\gamma=0$, the purely dissipative dynamics $\mathcal{D}\left( \rho_{dis} \right)=0$ drive the system towards a steady state $\rho_{dis}= \bigotimes \lvert\downarrow_x\rangle\langle\downarrow_x\rvert$ which does not commute with the Hamiltonian and thus ordered phases of the Hamiltonian can be frustrated by the dissipation. Moreover, in the specific case of $h^x/\gamma=0$, this Liouvillian belongs to a family of efficiently solvable dissipative models \cite{foss2017solvable} (see Appendix \ref{appendix:exact_solution} for further details) in which correlations remain localized and therefore the Liouvillian admits an efficient exact solution for local observables. We denote this method \textit{EXACT} and use it as a benchmark.

\begin{figure*}[!ht]
  	\includegraphics[width=\textwidth]{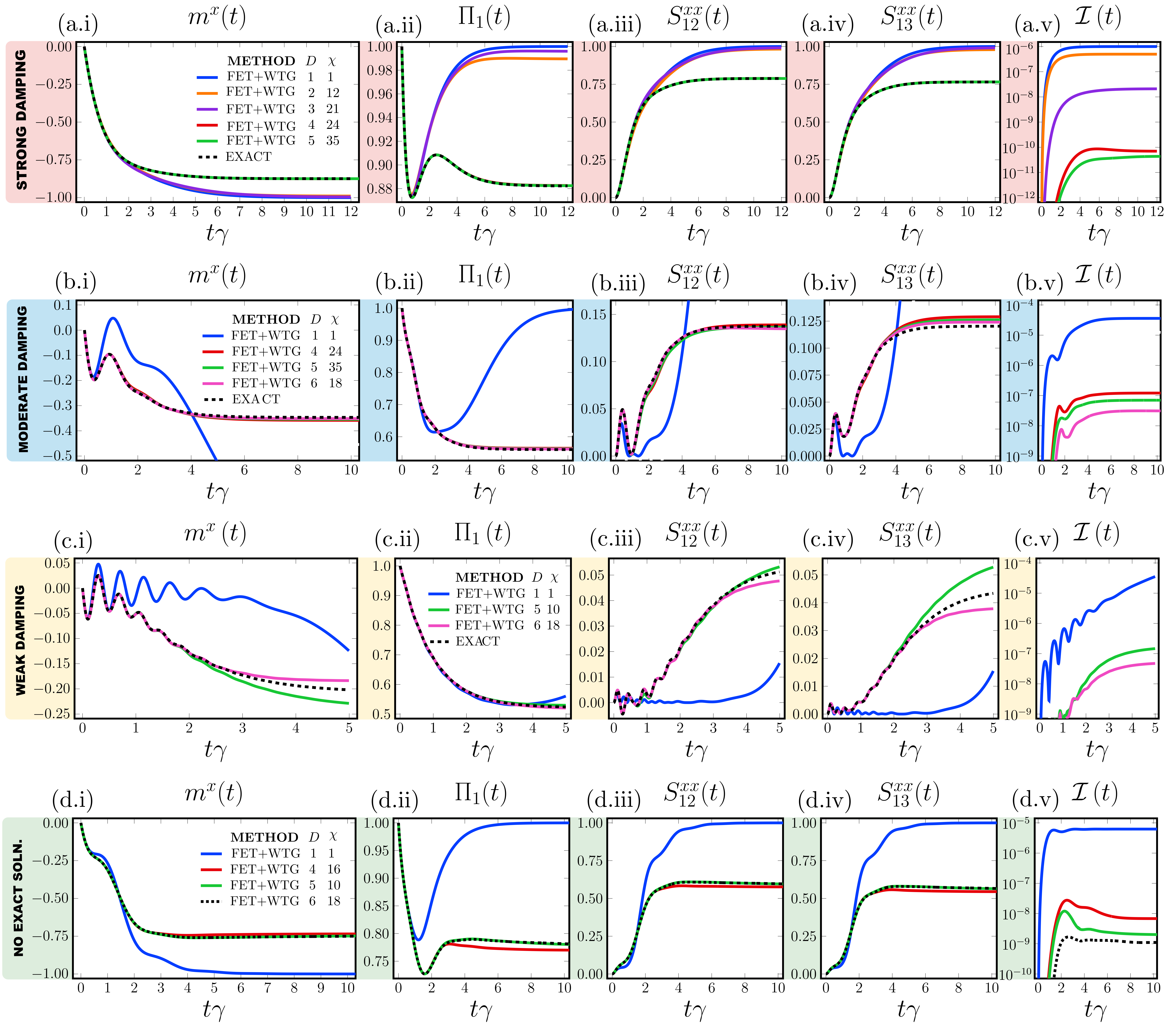}
  	\caption{\label{fig:main_results} Dynamics of the dissipative Ising model for $h_x/\gamma=0$ in $(a)$ strong ($V/\gamma=0.2$), $(b)$ moderate  ($V/\gamma=1.2$) and  $(c)$ weak ($V/\gamma=4.0$) spin damping regimes calculated using \METHOD for a range of bond dimensions $D$ and superimposed with results calculated using the EXACT method. $(d)$ Shows results for a regime not applicable to the EXACT method ($V/\gamma=0.5$ and $h_x/\gamma=1.0$) but which can be treated with \METHOD. In each case we plot (\romannumeral 1) the magnetization $m^{x}(t)$, (\romannumeral 2) the average purity $\Pi_1$ of the single site reduced density matrices, (\romannumeral 3) the nearest-neighbour $S^{xx}_{12}$, (\romannumeral 4) the next-nearest neighbour $S^{xx}_{13}$ spin-spin correlations and (\romannumeral 5) the average infidelity of truncation $\mathcal{I}(t)$ at each time step.}
\end{figure*}

For all parameters considered, we initialize the lattice spins in a product state $\rho_0 = \bigotimes\lvert\uparrow_z\rangle\langle\uparrow_z\rvert$ and simulate their evolution in time in strongly dissipative ($V/\gamma=0.2,\,h^x/\gamma=0$), moderately dissipative ($V/\gamma=1.2,\,h^x/\gamma=1.0$) and weakly dissipative ($V/\gamma=4.0,\,h^x/\gamma=0$) regimes, as well as in a regime ($V/\gamma=0.5,\,h^x/\gamma=1.0$) which does not admit an efficient solution using the EXACT method. For all results pertaining to this model we choose $\epsilon_{D'}=10^{-8}$ and set the convergence criteria for both the CTMRG and FET algorithms to $10^{-10}$. We choose a time step $\tau\gamma=0.01$ in all cases except for the weakly dissipative regime where we choose $\tau\gamma=0.005$.

In each regime we calculate reduced density matrices $\rho_j$ and $\rho_l$ for each lattice site labelled $j$ and $l$ in the two-site unit cell as well as the set of four nearest neighbour reduced density matrices $\rho_{jl}$ and four next nearest neighbour reduced density matrices $\rho_{jj'}$ where $j$ and $j'$ are at a distance of 2 lattice constants rather than $\sqrt{2}$, i.e. they are in the same row or column. Although we find that all reduced density matrices within each set are equivalent to a high precision, it is convenient to plot expectation values averaged over each set. We therefore calculate the average magnetization $m^x = \frac{1}{2}\left( \textrm{tr}\left(\hat\sigma^x\hat\rho_j\right) + \textrm{tr}\left(\hat\sigma^x\hat\rho_l\right)\right)$  as well as the average purity of the single site reduced density matrices $\Pi_1 = \frac{1}{2}\left( \textrm{tr}\left(\hat\rho_j^2 \right) + \textrm{tr}\left(\hat\rho_l^2 \right) \right)$ as function of time. To compare larger reduced density matrices we calculate $S^{xx}_{12}$ and $S^{xx}_{13}$, where $S_{jl}^{xx}(t)=\textrm{tr}(\hat\sigma^x_j\otimes\hat\sigma^x_l\rho^t)$, again averaged over the four possible choices for $j$ and $l$. Finally we show the infidelity $\mathcal{I}(t)=1-\mathcal{F}(t)$ of each truncation averaged over the four trotter layers which make up every time step $\tau$ where $\mathcal{F}$ is the mixed state fidelity equation (\ref{eqn:fidelity}). Results are plotted for a range of bond dimensions $D$ and the environment dimensions $\chi^{tr}$ and $\chi^{hs}$, where we choose $\chi^{tr}=\chi^{hs}=\chi$ in each case, and where $\chi^{tr}$ and $\chi^{hs}$ are associated to the effective environments $\mathcal{E}^{tr}$ and $\mathcal{E}^{hs}$, respectively. Finally, we have confirmed the convergence of the results with respect to increasing $\chi^{tr}$ and $\chi^{hs}$ in all results shown.

\subsubsection{Strong Dissipation}

In Fig. \ref{fig:main_results} (a) we plot the results of the strongly dissipative regime, in which the dissipative process dominates and where the spins are strongly damped. The exact dynamics of the system can be summarised as follows. From the initial product state, the average single site expectation value $\textrm{tr}\left( \sigma^x \rho_t \right)$ decays monotonically in time towards a steady state which reflects the strong spin damping. Each spin is initially in a pure state with $\textrm{tr}\left( \rho_t^2\right)=1$ and becomes mixed during the dynamics, eventually tending towards a purity of $\textrm{tr}\left( \rho_t^2\right)\approx0.88$ after the transient evolution. From an initially uncorrelated state, spin-spin correlations become non-zero and remain finite after the transient phase.

Comparing the results of \METHOD with the exact solution we find that excellent convergence is achieved for $D=4$ and $D=5$ while the results for $D=2$ and $D=3$ fall somewhere between the ``mean field''  $D=1$ solution and the exact solution. The $D=1$ solution tends towards an uncorrelated product state of spins in the $\lvert \downarrow^x\rangle$ phase which again reflects the dominance of the dissipative dynamics in the solution of the mean field theory. As correlations are included by increasing $D$ to $D=2$ and $D=3$ we find that $S^{xx}_{12}$ and $S^{xx}_{13}$ become non-zero and for $D=3$ the solution follows the exact dynamics closely at early times, however after the transient stage the spins tend towards an almost pure steady state in the $\lvert \downarrow^x\rangle$ phase, similar in character to the $D=1$ solution. Upon increasing to $D=4$ and $D=5$ we see that the \METHOD method reproduces the exact dynamics to excellent precision across all observables calculated.

Figure \ref{fig:main_results}  (a.\romannumeral 5) plots the infidelity of truncation $\mathcal{I}(t)$, the qualitative behaviour of which is similar for all values of $D$. As the the dynamics progress from the initial product state and correlations begin to deviate from zero, $\mathcal{I}(t)$ increases from $\mathcal{I}\ll1$ where the error introduced by truncation of enlarged bonds is negligible, to a larger finite value which indicates that the truncation causes the state to deviate slightly from the exact dynamics, nevertheless, for $D=4$ and $D=5$, $\mathcal{I}(t)$ remains below $\approx 10^{-10}$ at all times and is an indicator of the accuracy of the results. We note here that, $\mathcal{I}(t)$ has a dependence on the time step $\tau$ and this should be considered when comparing this parameter across different values of $\tau$. 

\subsubsection{Moderate and Weak Dissipation}
An example of the moderate dissipation regime is presented in Fig. \ref{fig:main_results} (b). In this case, the hopping strength is comparable to the dissipation and therefore the exact dynamics display some transient oscillations which are quickly damped by the dissipation. Here again, the exact solution contrasts significantly from the $D=1$ solution in which the dynamics tend towards a pure steady state with all spins in the $\lvert\downarrow\rangle$ state. We find that \METHOD reproduces the exact dynamics to good precision for the single site observables for $D>3$. While $S^{xx}_{12}$ and $S^{xx}_{13}$ also show good agreement with EXACT.

A weak dissipation case for $V/\gamma=4.0$ and $h^x/\gamma=0.0$ is plotted in Fig. \ref{fig:main_results} (c). The weakly damped oscillations of the EXACT results at early times reflect the dominance of the hopping term in this regime. While the $D=1$ solution gives incorrect results, the results for $D=5$ and $D=6$ reproduce the exact solution early in the transient phase and begin to deviate from the exact dynamics after approximately $t\gamma=2-3$ while still retaining the same qualitative behaviour. The fact that a larger bond dimension is required to reproduce the exact results is indicative of the greater role played by correlations in this coherent hopping dominated regime.

\subsubsection{Outside Exactly Solvable Regime}
For finite transverse field $h^x$, the Lindblad master equation does not fulfil the conditions for an efficient exact solution using the EXACT method and correlations may not remain localised, nevertheless \METHOD makes no assumption as to extent of correlations and should therefore be applicable for these parameters. As an example, a case for $V/\gamma=0.5$ and $h^x/\gamma=1.0$ is presented in Fig. \ref{fig:main_results} (d). Using \METHOD we find that the dynamics converge as the iPEPO bond dimension is increased. Results for $D \in (1,4,5,6)$ converge very well for $D\geq5$. The behaviour of the system is similar to the efficiently solvable cases; after some transient phase, the initial pure product state tends towards a correlated mixed state which is qualitatively different from the mean field solution. The infidelity of truncation Fig. \ref{fig:main_results} (d.\romannumeral 5) remains below $\mathcal{I}(t)<10^{-8}$ for the converged results, which is in line with previous benchmarking results.

\subsection{Comparison with Simple Update}\label{subsection:comparison_with_simple_update}
To highlight differences between the \METHOD and \SU truncation methods, we compare the results calculated using each method in the moderate damping regime ($V/\gamma=1.2$, $h^x/\gamma=0$) of section \ref{subsection:IsingModel} for a range of bond dimensions. All parameters are the same for both methods; $\tau=0.01$ and $\epsilon_{D'}=10^{-8}$ and CTRMG and FET convergence criteria set to $10^{-10}$, with the only difference being in how $\tilde u$, $\tilde v$ and $\tilde\sigma$ are calculated. 

\begin{figure}[b!]
  	\includegraphics[width=\linewidth]{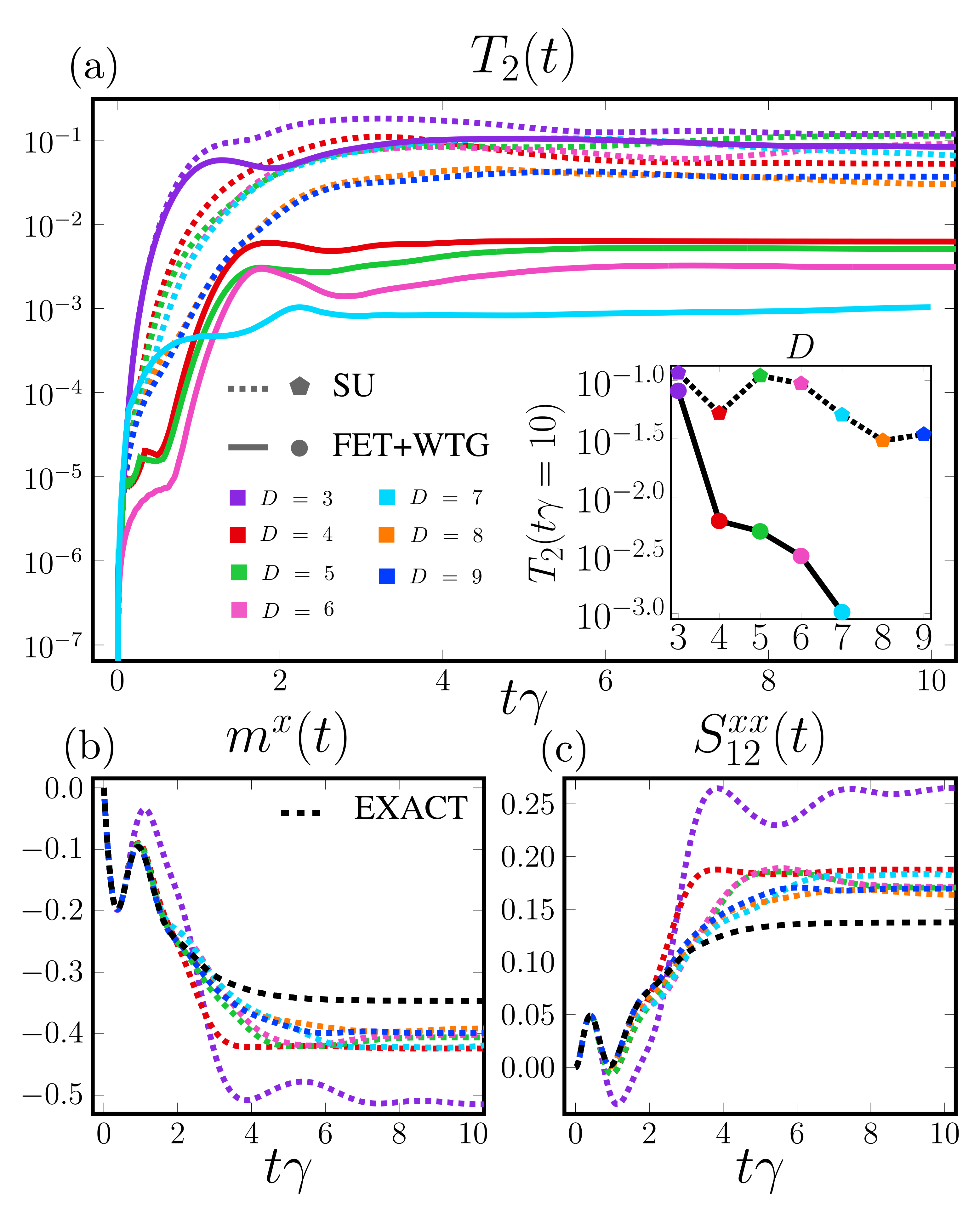}
  	\caption{\label{fig:SUCOMPARISON}Comparison between \METHOD (solid), \SU (dotted) and EXACT (dashed line) in the moderately damped regime of the dissipative Ising model $V/\gamma=1.2$, $h^x/\gamma=0.0$. $(a)$ Trace distance $T_{2}(t)$ as a function of time and at $t\gamma=10$ (inset) for a range of bond dimensions. $(b)$ Magnetization $m^{x}(t)$ and $(c)$ nearest-neighbour $S^{xx}_{12}$ show that \METHOD outperforms SU.}
\end{figure}

As well as comparing the observables $m^x(t)$ and $S^{xx}_{12}(t)$, we provide a quantitative measure of the accuracy of each method by calculating the trace distance between the EXACT reduced density matrix at each time step and the corresponding reduced density matrix calculated using the different TN methods. In particular we find the trace distance $T_{2}(t)$ of the nearest neighbour reduced density matrices $T_2\left(\rho_{jl},\phi_{jl}\right) = \frac{1}{2}\textrm{tr}\left( \sqrt{(\rho_{jl}-\phi_{jl})^{\dagger}(\rho_{jl}-\phi_{jl})}\right)$ where $T_{2}(t)$ is averaged over the four nearest neighbour reduced density matrices of the two site unit cell. By observing $m^x(t)$, and $S^{xx}_{12}$ and the trace distance $T_{2}$ in Fig. \ref{fig:SUCOMPARISON} (a-c) it is clear that the \SU method does not reproduce the EXACT results to the same accuracy as \METHOD. Fig. \ref{fig:SUCOMPARISON} (a) and its inset demonstrates that, while \METHOD shows clear systematic improvement in accuracy as $D$ is increased, \SU shows only minor and not clearly systematic reduction in $T_2(t\gamma=10)$ even if $D$ is increased well beyond that for which \METHOD demonstrates good convergence. For values of $D>3$, $T_{2}$ is consistently about an order of magnitude smaller for \METHOD than for SU, demonstrating the much better compression and greater accuracy of WTG+FET. The observables in Fig. \ref{fig:SUCOMPARISON} (b-c) calculated using \SU deviate from the EXACT dynamics considerably compared to \METHOD (compare to Fig. \ref{fig:main_results} (b)), at times $t\gamma\gtrapprox2$, the \SU method struggles to accurately capture the EXACT dynamics for all bond dimensions shown. 

\begin{figure}[t!]
  	\includegraphics[width=\linewidth]{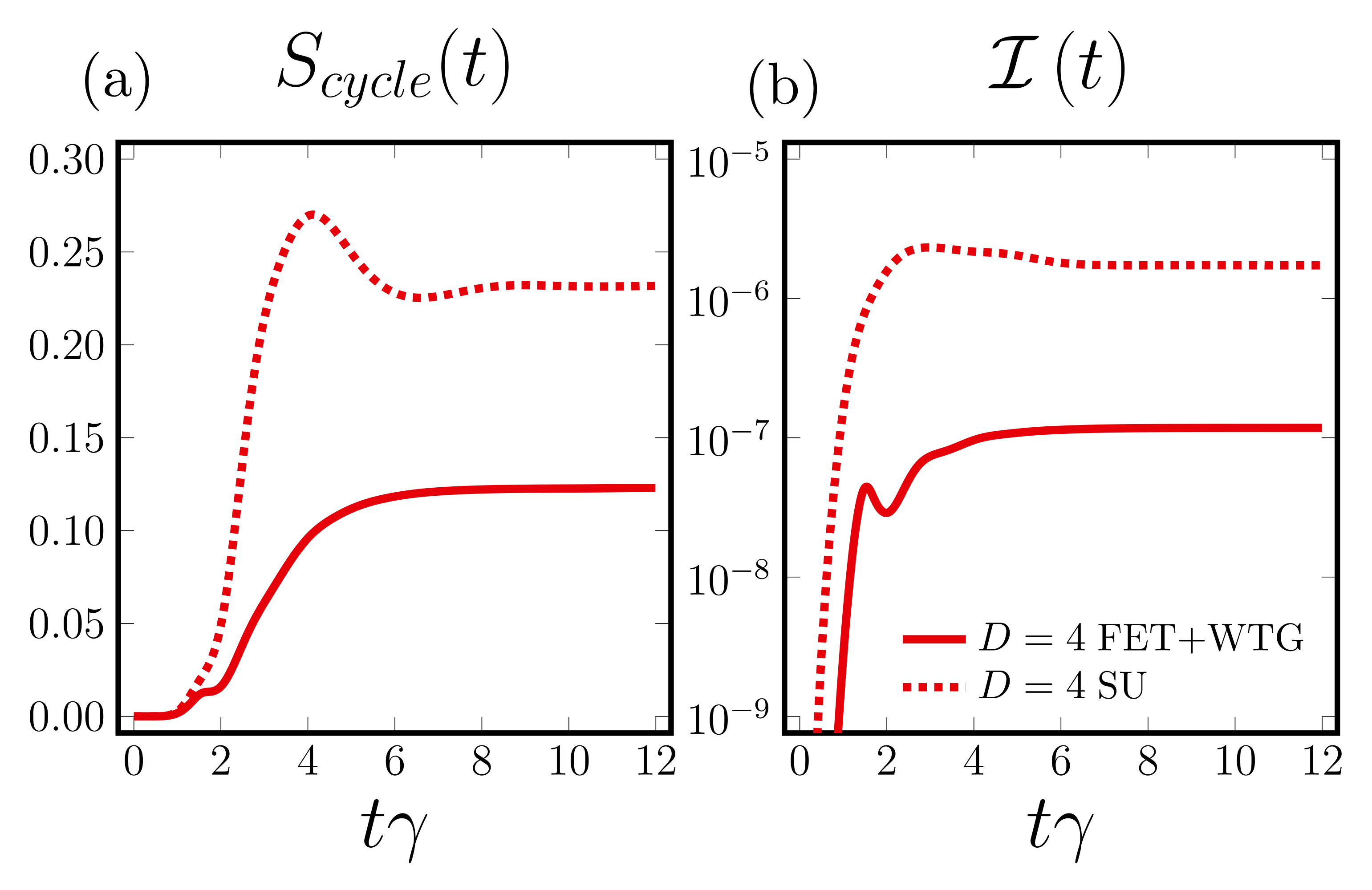}
  	\caption{\label{fig:CYCLE_ENTROPY} $(a)$ The accumulation of internal correlations in time quantified by the cycle entropy $S_{cycle}(t)$ is more effectively curtailed by \METHOD and $(b)$ the infidelity of truncation $\mathcal{I}(t)$ is an order of magnitude smaller than \SU at each truncation step. Results form moderately damped regime of the dissipative Ising model $V/\gamma=1.2$, $h^x/\gamma=0.0$ with $D=4$.}
\end{figure}

Finally we compare how the two algorithms deal with internal correlations in the network and compare the fidelity of truncation at each time step. TNs with closed loops (or cyclic TNs) can suffer from an accumulation of \textit{internal} correlations, which do not contribute to any property of the quantum state. To achieve an optimal TN representation of the state at each truncation step, it is necessary to remove these internal correlations. Furthermore, a build up of these correlations can lead to problems in computation and breakdown of algorithms \cite{evenbly2018gauge}. The cycle entropy $S_{cycle}$ defined in \cite{evenbly2018gauge} prescribes a way of quantifying the extent of internal correlations in the network, and is conveniently expressed in terms of the bond environment, details of its calculation in the present case are given in Appendix \ref{appendix:cycle_entropy}. The cycle-entropy $S_{cycle}$ plotted in Fig. \ref{fig:CYCLE_ENTROPY} (a) shows the extent of internal correlations in the network as a function of time. Initially the network, which represents a product state, has no internal correlations. In time, the extent of internal correlations grows and saturates at a finite value. Importantly, $S_{cycle}$ grows more slowly and saturates at a smaller value for \METHOD than it does for SU, illustrating that the proper truncation of bonds reduces the extent of internal correlations in the network. Although the growth of $S_{cycle}$ in this case is relatively benign, the failure of SU to curtail the accumulation of internal correlations may contribute to the breakdown of the algorithm in some circumstances. As a final comparison we plot the infidelity of truncation $\mathcal{I}$ as a function of time for the two different methods in Fig. \ref{fig:CYCLE_ENTROPY} (b) and find that the \METHOD method outperforms \SU, decreasing the infidelity between truncated an untruncated bonds by approximately an order of magnitude. Although the variational degree of the ansatz is the same in each case --- they have same $D$ and $\chi$ --- the method by which enlarged bonds are truncated is crucially important in finding an optimal representation, thereby greatly reducing the accumulation of errors do to inadequate truncation and ultimately giving the most accurate results.

\subsection{A Driven-Dissipative Hard Code Boson Model}\label{subsection:drive_dissipative_XY_model}
In driven-dissipative quantum lattice models dissipation to the bath is replenished via a coherent or incoherent drive. Driven-dissipative systems constitute an important class of models with direct relevance to experimental platforms such as driven coupled photon arrays in a variety of architectures \cite{carusotto2013quantum}. In this section we calculate steady state properties of a driven-dissipative hard core boson model which can be mapped to a lattice of interacting spin-1/2 particles. The Hamiltonian is given in the rotating frame by
\begin{equation}
\label{eqn:XYHamiltonian}
\hat H = \sum_{j} \left[ -\Delta \hat \sigma^{+}_j \hat \sigma^{-}_j + F\left(\hat \sigma^{+}_j + \hat \sigma^-_j\right) \right]
-\frac{\textrm{J}}{z} \sum_{\langle j,l\rangle} \hat \sigma^{+}_j \hat \sigma^{-}_ l,
\end{equation}
where $\Delta = \omega_{p}-\omega_{c}$ is the detuning between the pump frequency $\omega_p$ and the on site energy $\omega_{c}$, $F$ is the pump field strength, $J$ is the hopping coupling and the sum $\sum_{\langle j,l\rangle}$ runs over nearest neighbours in the lattice of coordination number $z$. The spins undergo dissipation at a rate $\gamma$ described by a Lindblad operator $\hat L_j = \sqrt{\gamma}\hat \sigma ^{-}_j$, which is the same at each site and where the spin raising and lowering operators are defined as $\hat\sigma^{\pm} \equiv \frac{1}{2}(\hat\sigma^x\pm i\hat\sigma^{y})$.

\begin{table}[t!]
\centering
\caption{Steady state values of a hard core boson model on an infinite square lattice with parameters $\Delta/\gamma=5.0$, $F/\gamma=2.0$ and $J/\gamma=1.0$ calculated using \METHOD. In each case we use a time step of $\tau\gamma=0.0025$. For comparison we tabulate results for the same parameters from the corner space renormalization method \cite{finazzi2015corner} for different sizes $N_x\times N_y$.}
\renewcommand{\arraystretch}{1.5}
\begin{tabular}[t]{llllll}
\multicolumn{6}{c}{$J/\gamma=1.0 \quad F/\gamma=2.0 \quad \Delta/\gamma=5.0$}\\
\hline\hline
$D\;$ &$\chi\quad$ & $\epsilon_{D'}\quad\quad$ & $n\quad\quad\quad$ & $\Re(\langle \hat \sigma^{-} \rangle)\quad\quad$ & $g^{(2)}_{\langle j,l\rangle}\quad\quad$ \\
\hline
\multirow{1}{*}{1}  & 1 	& $10^{-6}$		& 0.09482   & 0.27619     	& 1.0     \\\cline{1-6}
\multirow{5}{*}{3}  & 9 	& $10^{-4}$		& 0.09545  	& 0.27674    	& 1.06243 \\
					& 9 	& $10^{-5}$		& 0.09534  	& 0.27680    	& 1.06353 \\
					& 9 	& $10^{-6}$		& 0.09534   & 0.27681	    & 1.06360 \\
					& 9 	& $10^{-7}$		& 0.09535   & 0.27680	    & 1.06344 \\
					& 15 	& $10^{-7}$		& 0.09535  	& 0.27680    	& 1.06344 \\\cline{1-6}
\multirow{2}{*}{4}  & 8  	& $10^{-7}$		& 0.09548  	& 0.27670       & 1.06440 \\
 		     		& 12 	& $10^{-7}$		& 0.09548   & 0.27670       & 1.06443 \\\cline{1-6}
\multirow{2}{*}{5}  & 10 	& $10^{-7}$		& 0.09548  	& 0.27670       & 1.06443 \\
 		     		& 15 	& $10^{-7}$		& 0.09548   & 0.27670       & 1.06443 \\\cline{2-6}
\hline\hline
\multicolumn{6}{l}{$N_x\times N_y\quad$ Corner Space Renormalization Method}       		\\\hline
\multicolumn{2}{l}{$4\times4$}  & 			& 0.0954(1) 	& 0.2764(2) & 1.0643(3)   	\\
\multicolumn{2}{l}{$8\times4$}	& 			& 0.09527(2)	& -         & 1.0436(3)     \\
\multicolumn{2}{l}{$8\times8$}	& 			& 0.0948(2) 	& -         & 1.0237(6)     \\
\hline\hline
\label{tab:HCB}
\end{tabular}
\end{table}

We compare steady state expectation values with those calculated using the Corner Space Renormalization method \cite{finazzi2015corner}. To this end we consider an array of hard core bosons with $\Delta/\gamma=5$, $F/\gamma=2$ and $J/\gamma=1$ and calculate the average single site boson density $n=1/2(n_j+n_l)$, the nearest neighbour ($\langle j,l\rangle$)
correlation functions $g^{(2)}$ averaged over all combinations of ($\langle j,l\rangle$), where
\begin{equation}
g^{(2)}_{j,l} = \frac{\langle\hat \sigma^{+}_j\hat \sigma^{+}_l\hat \sigma^{-}_j \hat \sigma^{-}_l\rangle}{ \langle\hat \sigma^{+}_j\hat \sigma^{-}_j\rangle\langle\hat \sigma^{+}_l\hat \sigma^{-}_l\rangle }, \quad n_j = \textrm{tr}(\hat \sigma^{+}_j\hat \sigma^{-}_j \rho_{ss}).
\end{equation}
Finally we calculate the average real part of $\Re\left[ \textrm{tr}\left(\hat \sigma^{-}  \rho_{ss} \right)\right]$ at each lattice site.

Staring from  an initial product state, we find the steady state for a set of parameters $D$, $\chi$ and $\epsilon_{D'}$, where convergence in time is achieved when all expectation values $\hat o$ up to next nearest neighbour fulfil a convergence criterion of $\epsilon_t < 10^{-6}$ where
\begin{equation}
\epsilon_t = \frac{\lvert\textrm{tr}\left(\hat o \rho_{t+\tau} \right) - \textrm{tr}\left(\hat o \rho_{t} \right)\rvert}{\lvert\textrm{tr}\left(\hat o \rho_t \right)\rvert\tau}.
\end{equation}
We use the steady state iPEPO calculated for one set of variational parameters as an initial state for the next until convergence to the desired precision is achieved. Results of this procedure are given in TABLE \ref{tab:HCB} along with comparable results from \cite{finazzi2015corner}. 

The steady state values converge as the iPEPO variational parameters are increased and are comparable to the results of the Corner Space Renormalization method. Where we might expect increasing the Corner Space Renormalization lattice size $N_x\times N_y$ will give results closer to the \METHOD method, which represents the thermodynamic limit directly, we find that the opposite is true, with a lattice size of $4\times 4$ closer in agreement to \METHOD than $8 \times 8$. This discrepancy could be due to finite size effects or spatial symmetry breaking, which may be present in the $N_x\times N_y$ results, and is not observed in the iPEPO solution where we have enforced two-site translational invariance by choosing a two-site unit cell.

\subsection{Anisotropic Dissipative XY Model}\label{subsection:anisotropic_dissipative_XY_model}
\begin{figure}[t!]
  	\includegraphics[width=\linewidth]{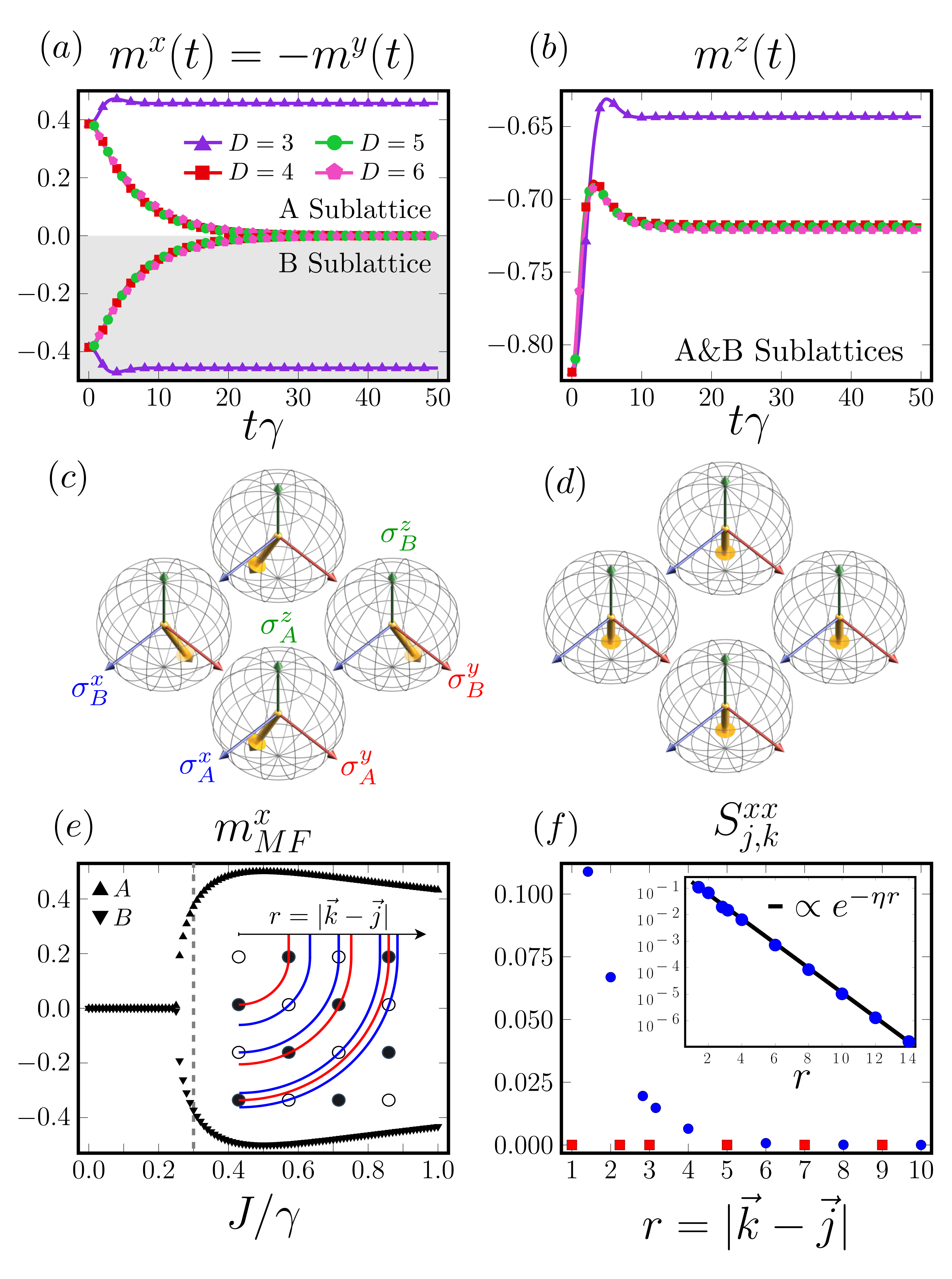}
  	\caption{\label{fig:sXY} \textbf{Fate of staggered-XY phase at $\bm{J/\Gamma=0.3}$.} (a-b) Local magnetizations $m^{x,y,z}(t)$ on the A and B sublattices  as the state evolves from the $D=1$ steady state solution for bond dimensions $D\in[3,4,5,6]$. (c-d) Representation of a $2\times2$ plaquette of the lattice in (c) the staggered-XY phase ($D=1$ steady state) and (d) the uniform phase ($D=6$ steady state). (e) Mean field phase diagram with transition at $J/\Gamma=\frac{1}{4}$. Inset: Radii $r=\lvert \vec{k}-\vec{j}\rvert$ of an odd (red) and even (blue) number of steps on the lattice. (f) The correlation function $S^{xx}_{j,k} = \langle \sigma^x_j\sigma^x_k\rangle$ versus distance $r$ has a staggered form which is a remnant of the staggered-XY phase; correlations at odd step radii (red squares) are  zero and those at even step radii (blue circles) are finite and decaying with $r$. Inset: Exponential fit to even step correlations giving exponent $\eta\approx1.07$.}
\end{figure}

Having demonstrated the capabilities of the algorithm, by choosing an interesting example we now show that the method is highly suitable to address  physical questions. In particular, two dimensional systems can host a unique set of phenomena, here we explore the stability with respect to fluctuations of a spontaneously symmetry broken staggered-XY (sXY) phase in the steady state of an ansiotropic dissipative XY model. While the mean field theory predicts that the sXY phase is stable in two dimensions, it is not clear whether it remains accessible if fluctuations at the microscopic level are accounted for and if any long range order associated with the sXY phase is present. The anisotropic dissipative XY model has a Hamiltonian of the form
\begin{equation}
\label{eqn:AnisotropicXYHamiltonian}
\hat{H} = \frac{J}{z}\sum_{\langle j,k\rangle}{\hat{\sigma}^x_j\hat{\sigma}^x_k - \hat{\sigma}^y_j\hat{\sigma}^y_k},
\end{equation}
with a nearest neighbour hopping $J$ and coordination number $z=4$, as well as dissipation described by local Lindblad operators $\hat{L}_j=\sqrt{\Gamma}\hat{\sigma}_j^-$ at each lattice site. The (Gutzwiller) mean-field (MF) phase diagram, plotted in Fig. \ref{fig:sXY}. (e), was studied in \cite{PhysRevLett.110.257204} and shows that, for $J/\Gamma>1/4$, the steady state hosts a staggered-XY symmetry broken phase in which the spins divide into A and B sublattices with angles $\pm\theta$ relative to the $x=y$ line on the Bloch sphere as depicted in Fig. \ref{fig:sXY}(c). The spontaneous breaking of this continuous $U(1)$ symmetry means that $\theta$ can take any value and allows for vortexlike topological defects in the lattice. The question of whether or not the sXY phase is accessible in two dimensions if corrections beyond MF theory are accounted for has previously been addressed using a Keldysh field theory approach \cite{PhysRevLett.110.257204,PhysRevB.93.014307}. There, an effective model is constructed by mapping the spins to bosons, an approach which does not capture the microscopic physics of the spin model but addresses the behaviour in the long wavelength limit. In that approximation they found that the steady state physics of the effective model is described by a partition function in the same universality class as the classical XY model and therefore one should expect a Kosterlitz Thouless transition in two dimensions. However it is also predicted in \cite{PhysRevB.93.014307}, based on a simple MF theory analysis, that the effective temperature of the model will be greater than the Kosterlitz Thouless temperature, such that the ordered phase will not be accessible when quantum fluctuations are included and any long range algebraic order will be absent or at least significantly diminished. We can now use our method to address this question exactly by directly solving the microscopic spin model close to the transition point $J/\Gamma=1/4$, where the MF theory is expected to break down. Moreover, we are able to give a quantitative picture of the system by calculating not only local observables as a function of time, but also spatial correlation functions in the steady state.

We first find the steady state iPEPO representation of the model for a bond dimension $D=1$ --- equivalent to a MF solution --- at $J/\Gamma=0.3$, which lies just within the sXY phase. To do this, we initialize the iPEPO in a state for which the symmetry is explicitly broken $\langle\sigma^x_A\rangle=-\langle\sigma^x_B\rangle=1$ and calculate the $D=1$ steady state with WTG+FET. Then, using the symmetry broken $D=1$ iPEPO solution as an initial state, we systematically add quantum fluctuations by calculating steady states for bond dimensions $D\in[3,4,5,6]$ until convergence. Results of this procedure are presented in Fig. \ref{fig:sXY}.
For bond dimensions $D=3$ we find that the system remains in the sXY phase. For $D\in[4,5,6]$, however, the spin magnetization $m^z(t)$, which is uniform across the lattice, is slightly modified and the magnetizations $m^x(t)$ and $m^y(t)$ on each sublattice slowly tend towards zero such that the continuous symmetry is no longer broken---depicted in Fig. \ref{fig:sXY} (d)---and the sXY phase is therefore unstable to fluctuations, corroborating the Keldysh field theory predictions of \cite{PhysRevB.93.014307}. This proves that long wavelength fluctuations captured by the approximate theory dominate over other microscopic fluctuations.
In Fig. \ref{fig:sXY} (f) we plot the correlation function $S^{xx}_{k,j}=\langle\sigma^x_j\sigma^x_k\rangle$ (note that $\langle\sigma^x_j\rangle\langle\sigma^x_k\rangle=0$) which shows a staggered structure reminiscent of the sXY phase where correlations at a radii $r$ (see Fig. \ref{fig:sXY} (e) inset) corresponding to an odd number of steps on the lattice are zero, whereas even step correlations are finite and decay with $r$. Considering only the even step correlations in Fig. \ref{fig:sXY}. (f) inset, we find that the decay is well approximated by an exponential function of the form $S^{xx}_{r\in \textrm{even}}\propto e^{-\eta r}$ with $\eta \approx 1.07$, any long range algebraic order which may have been associated to the symmetry broken phase is not present in the iPEPO solution suggesting that the system is in the disordered phase. Good convergence is found for $D=6$ and $\tau\gamma=0.01$ resulting in infidelity of truncation $\mathcal{I}(t)<10^{-9}$.

\section{Discussion}\label{section:perspectives}
We have developed a new TN algorithm capable of accurately simulating dynamics of dissipative quantum lattice models on a two-dimensional square lattice directly in the thermodynamic limit. The method adapts the Full Environment Truncation (FET) and Weighted Trace Gauge (WTG) fixing techniques of \cite{evenbly2018gauge} to dealing with the iPEPO TN ansatz for mixed states. Comparisons with exact numerical results demonstrate an excellent accuracy of the method and its performance across different dissipative regimes. 
Contrasting with the more efficient but much less accurate simple update truncation scheme, we have proven that it is necessary to optimally truncate enlarged bonds to obtain accurate results. We have shown the applicability of the technique for calculating steady state properties of driven-dissipative systems by comparison with literature results. The methods performs well in regimes where mean-field approximation fails, proving able to capture substantial correlations in the presence of dissipation. Finally we have shown that a staggered-XY phase of the dissipative anisotropic XY model predicted by mean field theory is not stable if correlations are included and while a remnant of the staggered structure remains in the correlation function, it's decay is well approximated by an exponential function and no long range order remains. 

As with similar algorithms for iPEPS, the principal contribution to the computational complexity of the algorithm comes from the calculation of the effective environment which is updated at each time step (here using CTMRG). The leading cost of the version of CTMRG we use arises from a singular value decomposition of order $O(\chi_{hs}^3D^6)$, improvements in performance can therefore be achieved by optimizing this step, for instance, using a fixed point method such as the FPCM \cite{fishman2018faster} or approximating the effective environment by using a boundary matrix product state to represent the boundary of the system. Numerous algorithms have been developed to calculate the fixed point including a time-evolving block decimation (TEBD) \cite{orus2008infinite,vidal2003efficient} or variational MPS-tangent space methods (VUMPS) \cite{zauner2018variational,haegeman2016unifying,vanderstraeten2019tangent,fishman2018faster,nietner2020efficient} and can lead to significant speed up for TNs which are close to being critical \cite{fishman2018faster}.

As well as accurately determining steady state properties such as long range equal-time correlation functions, this work facilitates the calculation of more complex dynamical properties e.g. dynamical correlation functions and fluorescence spectra of strongly correlated driven dissipative quantum lattice models. A significant advantage of both the FET method of truncating enlarged bonds and the WTG method of fixing the TN gauge, is that they can be used in tensors networks of arbitrary geometries, provided the bond environment can be calculated efficiently. In this regard, straightforward adaptations of the method we have presented in this work could be used to treat driven-dissipative models with longer range interactions or those defined on more complicated network structures such as hyperbolic lattices \cite{kollar2019hyperbolic} as well as problems related to functional quantum biology \cite{lambert2013quantum,scholes2017using}.
\newpage


\appendix

\section{Calculating the Effective Environments}\label{appendix:calculating_the_environment}
\begin{figure}[!ht]
 \includegraphics[width=0.95\linewidth]{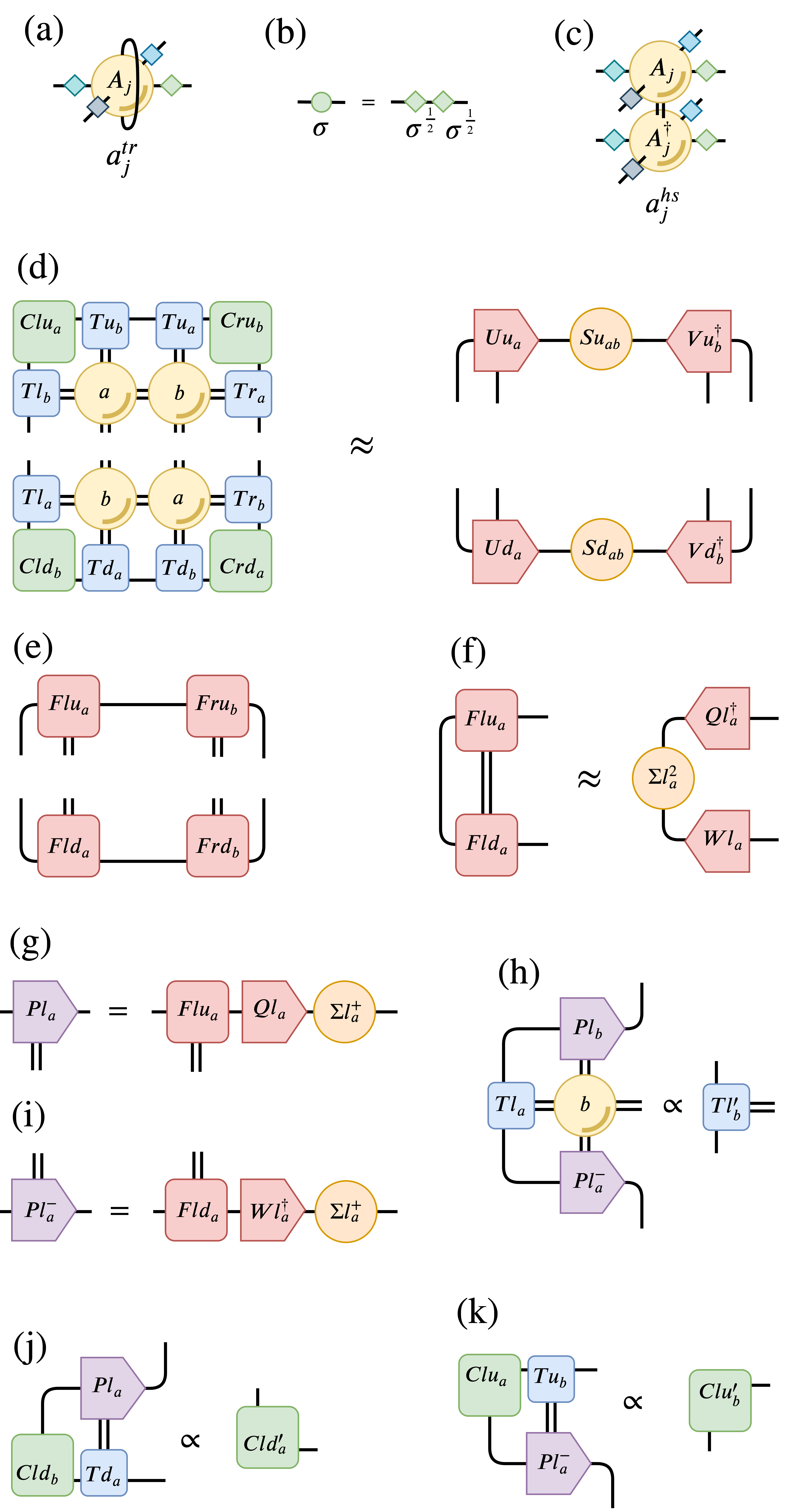}
  \caption{Tensor diagrams representing some of the steps involved in performing the \textit{left-move} component of the CTMRG algorithm used to calculate the effective environment $\mathcal{E}^{hs}$.}
  \label{fig:CTMRG}
\end{figure}

Given tensors representing the unit cell of the 2D lattice, we calculate the effective environments $\mathcal{E}^{tr}$ and $\mathcal{E}^{hs}$ using a variant of the Corner Transfer Matrix Renormalization Group (CTMRG) algorithm \cite{baxter1968dimers,baxter1978variational,nishino1996corner,nishino1997corner,orus2009simulation,corboz2010simulation,fishman2018faster}. To improve stability and convergence properties of the CTMRG algorithm as well as the conditioning of the bond environment $\Upsilon_{jl}$ we find it helpful to use the variant of CTMRG presented in \cite{fishman2018faster}, which makes use of an intermediate singular value decomposition. Following \cite{fishman2018faster} we refer to Fig. \ref{fig:CTMRG} in describing the basic steps involved in the \textit{left-move} component of the CTMRG algorithm used for calculating $\mathcal{E}^{hs}$ for an iPEPO with a two-site unit cell. 
This algorithm goes as follow: Consider the unvectorized sixth rank iPEPO tensors $A_j$ and $A_l$.
\begin{itemize}
\item (a) To calculate the trace effective environment $\mathcal{E}^{\textrm{tr}}$ we trace over the physical dimensions of the iPEPO unit cell tensors, giving the fourth rank tensors $a\rightarrow \textrm{tr}_{d}\left( A_j\right)$ and $b\rightarrow \textrm{tr}_{d}
\left( A_l\right)$, where we have split the bond environment two (Fig. \ref{fig:CTMRG} (b)) and contracted each half with $a$ and $b$ appropriately.
\item (c) Alternatively, to calculate the Hilbert-Schmidt effective environment $\mathcal{E}^{\textrm{hs}}$ we first find the Hilbert-Schmidt inner product over the physical indices of the vectorized $A_j$ and $A_l$ giving the eighth rank tensors $a\rightarrow \textrm{tr}_{d} (A_jA_j^{\dagger})$ and $b\rightarrow \textrm{tr}_{d}( A_lA_l^{\dagger})$. The \textit{left-move} CTMRG step then proceeds as follows, where the tensor diagrams of Fig. \ref{fig:CTMRG} show the eight rank versions of $a$ and $b$ and therefore represent steps in the calculation of $\mathcal{E}^{hs}$.
\item (d) We construct the upper and lower half system transfer matrices and take a SVD to find the upper and lower decompositions $Uu_a Su_{ab} Vu_b^{\dagger}$ and $Ud_a Sd_{ab} Vd_b^{\dagger}$.
\item (e) We define $Flu_a\equiv Uu_aSu_{ab}^{1/2}$, $Fru_a\equiv Su_{ab}^{1/2}Vu_a^{\dagger}$, $Fld_a\equiv Ud_aSd_{ab}^{1/2}$ and $Frd_a\equiv Sd_{ab}^{1/2}Vd_a^{\dagger}$ where singular values of magnitudes (relative to the largest singular value) less than some small tolerance are truncated to improve stability.
\item (f) We next use the so called \textit{biorthogonalization} procedure (see \cite{fishman2018faster} for further details) to calculate $Pl$ and $Pl^{-}$, the first step of which is to contract $Flu_a$ with $Fld_a$ and perform a SVD to find $Wl_a$, $Ql_a$ and the diagonal matrix $\Sigma l_a^2$.
\item (g,j) We calculate the projectors $Pl_a=Flu_a Ql_a \Sigma l_a^+$ and $Pl_a^- =Fld_a Wl_a^{\dagger} \Sigma l_a^+$ with $\Sigma l^+$  being the Moore-Penrose pseudoinverse of $\Sigma l$.
\item (f-h) We repeat steps (d-j) to calculate $Pl_b$ and $Pl_b^{-}$ by replacing $a\leftrightarrow b$ in the upper and lower half system transfer matrices. Using these projectors the updated environment tensors $Tl_b'$, $Tl_a'$, $Clu_a'$, $Clu_b'$, $Cld_a'$, $Cld_b'$ are calculated and normalized as shown in Fig. \ref{fig:CTMRG} (h,j,k). This is one iteration of the \textit{left-move} component of this CTMRG algorithm.
\end{itemize}

A similar sequence of steps is used to perform the \textit{right-move}, \textit{up-move} and \textit{down-move} steps in CTMRG. The set of directional moves are repeated in series until the vectors of singular values of the corner transfer matrices converge. It is possible to perform \textit{right-move} at the same time as \textit{left-move} by following the biorthogonalization routine starting with $Fru_b$ and $Frd_b$ calculated in step (b) above, similarly for \textit{up-move} and \textit{down-move}.

\section{Full Environment Truncation}\label{appendix:full_environment_truncation}
\begin{figure}[!ht] 
\includegraphics[width=0.9\linewidth]{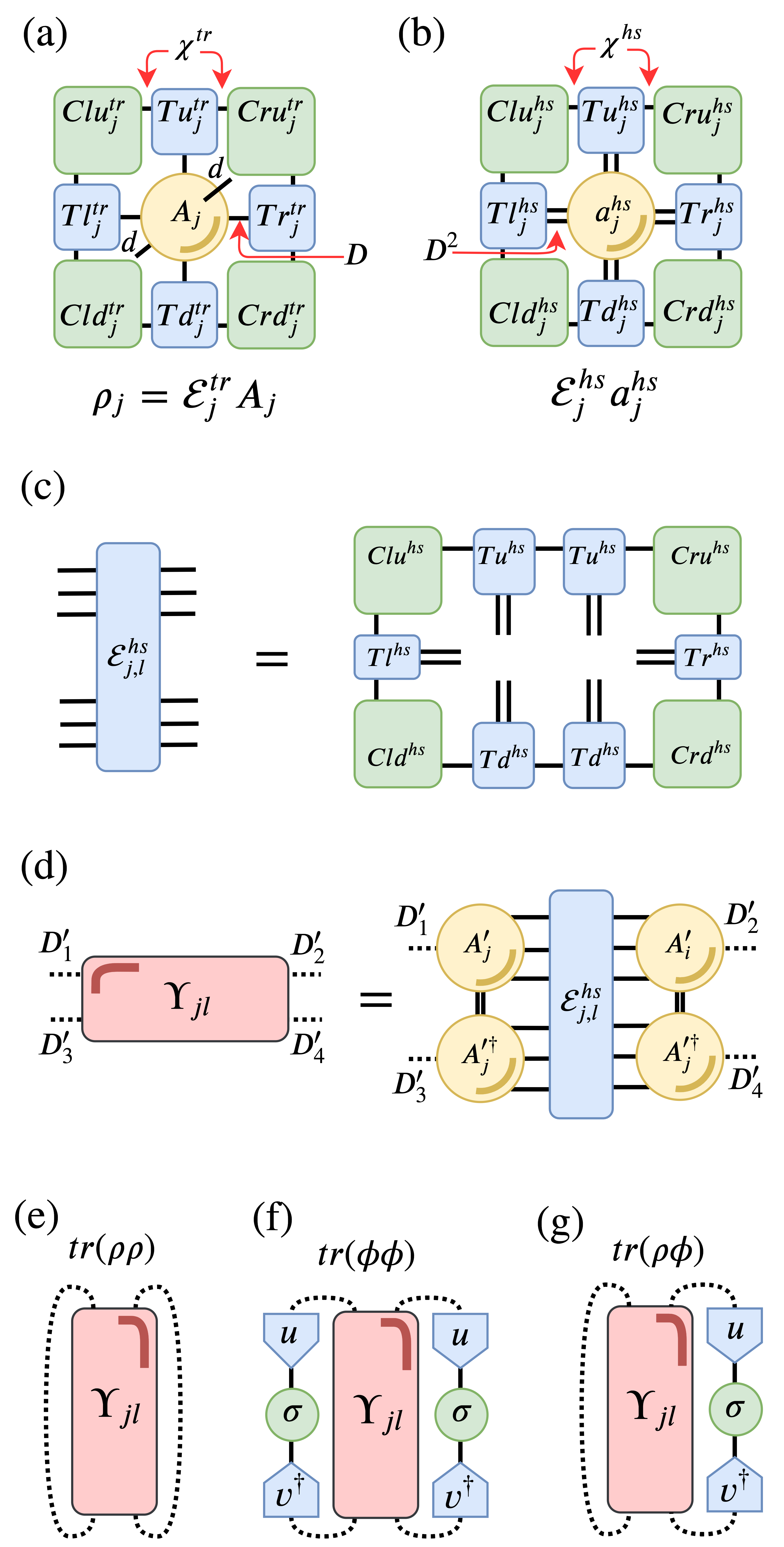}
\caption{The environment of the unit cell. (a) The trace effective environment $\mathcal{E}^{tr}_{j}$ of the iPEPO tensor $A_j$ used to calculate the $d\times d$ reduced density matrix $\rho_j$. (b) The Hilbert-Schmidt effective environment $\mathcal{E}_j^{hs}$ of the tensor $a_j^{hs}$ used in constructing the bond environment. (c) The effective environment $\mathcal{E}^{hs}_{j,l}$ of the tensors at neighbouring sites $j$ and $l$. (d) The bond environment $\Upsilon_{j,l}$ is the contraction of $\mathcal{E}_{j,l}^{hs}$ and the updated tensors $A'_j$ and $A'_l$ with enlarged bonds $\{D'_j\}\geq D$. (e-g) Using $\Upsilon_{j,l}$ the terms in the fidelity between the truncated ($\phi$) and untruncated ($\rho$) density matrices are calculated by contracting with the isometries $u$, $v$ and the bond matrix $\sigma$.}
\label{fig:bond_environment}
\end{figure}

An adapted Full Environment Truncation (FET) algorithm \cite{evenbly2018gauge} is used to truncate enlarge bonds of the iPEPO as follows. Let the state of the full system at time $t$ be $\rho_t$ and calculate the Hilbert-Schmidt environment $\mathcal{E}_{j,l}^{hs}$ of the iPEPO representing $\rho_t$ as discussed in section \ref{appendix:calculating_the_environment}. Find $A^{\prime}_j$ and $A^{\prime}_l$ by applying the Trotterized dynamical map and decompose the result via SVD retaining the $D'$ singular values with a magnitude (relative to the largest singular value) greater than $\epsilon_{D'}$. Contract $A^{\prime}_j$ and $A^{\prime}_l$ with the effective environment $\mathcal{E}^{hs}_{j,l}$ leaving only the enlarged bonds uncontracted as illustrated in Fig. \ref{fig:bond_environment} (d). This procedure leaves us with the fourth-rank bond environment tensor $\Upsilon_{jl}$.
Using the bond environment $\Upsilon_{ij}$, the tensors involved in the Rayleigh quotient proportional to $\mathcal{F}^2$ are calculated. Fig. \ref{fig:bond_environment} (e-g) illustrates the tensor contractions required to construct $\textrm{tr}\left(\rho\phi\right)$, $\textrm{tr}\left(\phi\phi\right)$ and $\textrm{tr}\left(\rho\rho\right)$ allowing us to represent $\mathcal{F}^2\left(\rho,\phi\right)tr\left(\rho\rho\right)$ in terms of the isometries $u$ and $v$, the bond matrix $\sigma$ and the bond environment $\Upsilon_{ij}$, where we note that the term $\textrm{tr}\left(\rho\rho\right)$ is independent of $u$, $\sigma$ and $v$.

\begin{figure}[h!]
  \includegraphics[width=\linewidth]{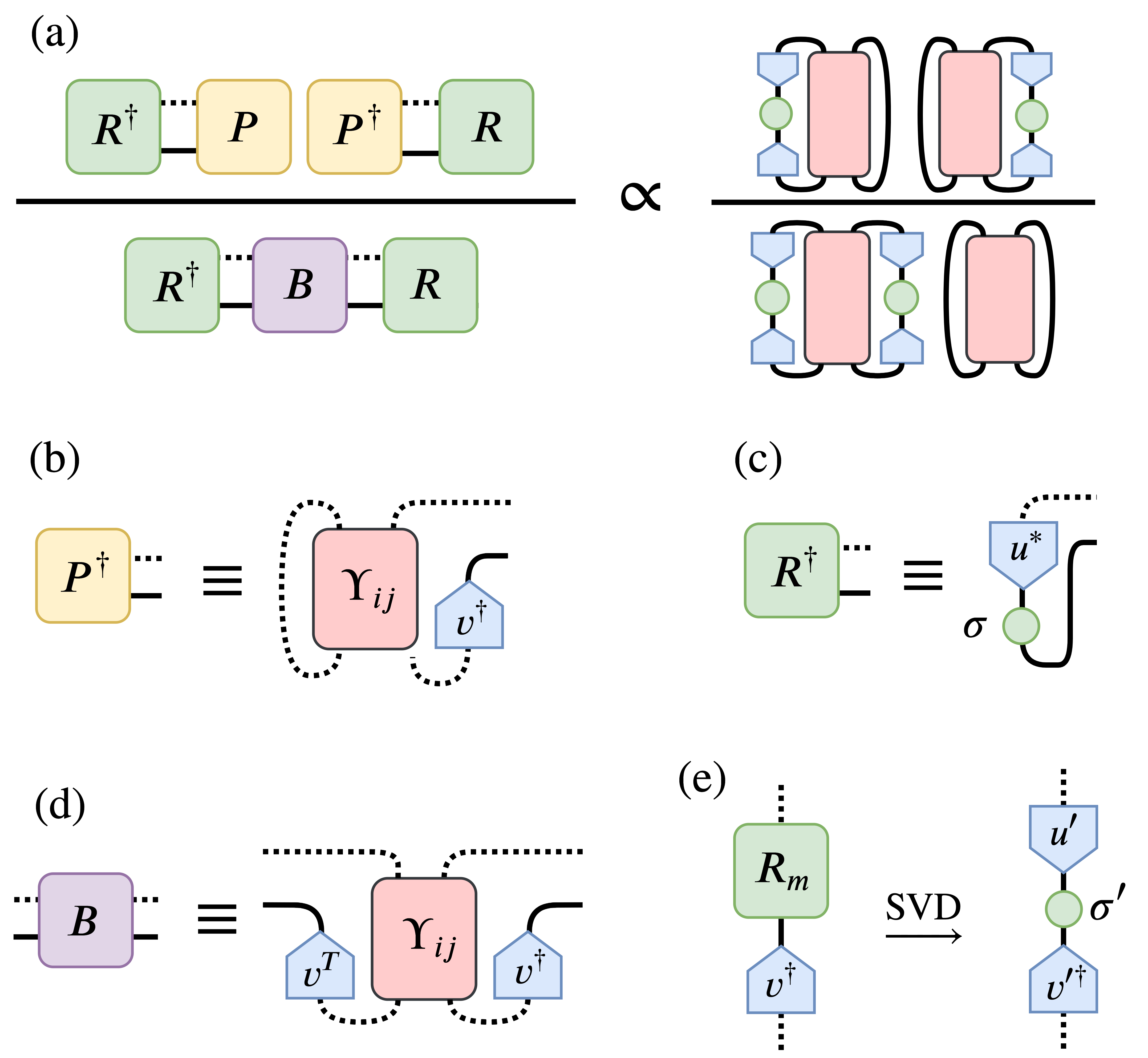}
  \caption{Tensor diagrams representing some of the steps involved in finding the isometries $\tilde u$ and $\tilde v$ and the bond matrix $\tilde \sigma$ which maximize the fidelity between the truncated and untruncated bonds. (a) The Rayleigh quotient in $R$ is proportional to $\mathcal{F}^2$. (b) $P$ is the contraction of the bond environment $\epsilon_{j,l}$ and the isometry $v$. (c) $R$ is the contraction of the bond matrix $\sigma$ and the isometry $u$. (d) $B$ is the contraction of $\Upsilon_{jl}$ with the isometry $v$. (e) The new (primed) isometries are found by singular value decomposition of the contraction of the maximal eigenvector $R_m$ and $v$.}
  \label{fig:FET_STEPS}
\end{figure}

The alternating optimization of $u$, $v$ and $\sigma$ proceeds as follows and is illustrated in Fig. \ref{fig:FET_STEPS}. Defining $R\equiv\sigma v$ (Fig. \ref{fig:FET_STEPS} (c)) the $R_m$ which maximizes $\mathcal{F}^2\left(\rho,\phi\right)\textrm{tr}\left(\rho^2\right)$ (Fig. \ref{fig:FET_STEPS} (a)) is found by keeping $v$ fixed and solving a generalized eigenvalue problem in $R$ (see Appedix \ref{appendix:optimizing_rayliegh_quotient} for further details). The updated tensors $\sigma'$ and $u'$ are then calculated using a SVD illustrated in Fig. \ref{fig:FET_STEPS} (e). Similarly by defining $L\equiv v'\sigma'$ the optimal $L_m$ is found giving $u''$, $\sigma''$ and $v''$. The alternating process is repeated until convergence of $\tilde{u}$, $\tilde{\sigma}$ and $\tilde{v}$ of isometries is reached.

\section{Optimizing Rayleigh Quotient}\label{appendix:optimizing_rayliegh_quotient}
A Rayleigh quotient of the form
$F(R) = \frac{\vec{R}^\dagger A \vec{R}}{\vec{R}^\dagger B \vec{R}}$
is maximized by the eigenvector $\vec{R}_{m}$, which corresponds to the largest eigenvalue $\lambda_{m}$ of the generalized eigenvalue problem $A\vec{R_i} = \lambda_iB\vec{R_i}$.
Since the matrix $A$ is constructed as an outer product $A=\vec{P}^{\dagger}\vec{P}$, the $\vec{R_m}$ which maximizes the Rayleigh quotient is given by $\vec{R_m}=\vec{P}B^{-1}$. In practice, it is possible to calculate $\vec{R_m}$ directly by inverting $B$ or by solving the system of linear equations $\vec{R_m}B=\vec{P}$ using, for example, a linear regression algorithm. Care must be taken at this stage to maintain the stability of the algorithm. If solving by direct inversion, we find it useful to either use a Moore-Penrose pseudoinverse \cite{penrose1955generalized} with some tolerance or by solving via linear regression with an intermediate truncated singular value decomposition. In our simulations we maximize the Rayleigh quotient by instead solving the generalized eigenvalue problem $A\vec{R}=\lambda B\vec{R}$ either by full diagonalization or by iterative methods to calculate only the maximal eigenvector $\vec{R_m}$ (Lancoz/Arnoldi).

\section{Exact Solution of Dissipative Ising Model}\label{appendix:exact_solution}
In order to provide a benchmark for our new TN method, we solve the dissipative transverse Ising model in section \ref{subsection:IsingModel} using the method of reference \cite{foss2017solvable}, which we briefly describe. As shown in \cite{foss2017solvable}, if a Liouvillian is structured such that coherences are not mapped to populations (and vice versa) then correlations in a system remain localized. This allows for an efficient exact determination of the time evolution of the local observables, which initially only have support on a suitably small sublattice.
In particular, an observable $O(t)$, which initially has support on a set of lattice sites $\mathcal{A}$, can be calculated at all times by solving in the Shrodinger picture:
\begin{equation}\label{eqn:exact}
O(t)=\textrm{Tr}_{\mathcal{A}\cup\mathcal{B}}\left[ \hat O \textrm{exp}\left(t\mathcal{L}_{\mathcal{A}\mathcal{B}} \right)\hat\rho_{\mathcal{A}\mathcal{B}}\right],
\end{equation}
where $\mathcal{B}$ is the set of lattice sites which are nearest neighbours of $\mathcal{A}$ and for which the Hamiltonian has simultaneous support on $\mathcal{A}$ and $\mathcal{B}$. 

We choose to calculate up to next nearest neighbour (in a lattice row or column) correlations $S^{xx}_{jl}(t)$ in time and therefore choose as $\mathcal{A}$ the set of three contiguous lattice sites in a row (in either the $x$ or $y$ lattice dimension) of the infinite two dimensional lattice. For a two-local Liouvillian, $\mathcal{B}$ is identified as the eight nearest-neighbour lattice sites of $\mathcal{A}$. Observables $O(t)$ can then be calculated efficiently by solving equation (\ref{eqn:exact}) using standard techniques from quantum optics (we used the Julia package QuantumOptics.jl \cite{kramer2018quantumoptics} to calculate the exact results).

\section{Cycle Entropy}\label{appendix:cycle_entropy}
For closed systems, $S_{cycle}$ is defined as the von-Neumann entropy of the normalized spectrum of a bond environment left contracted with the bond matrix $(\sigma\otimes\sigma)\Upsilon$ and is constructed as an inner product of pure states, (see \cite{evenbly2018gauge} for details). Here we instead use the bond environment left contracted with the bond matrix $(\sigma\otimes\sigma)\Upsilon$ which is constructed using $\mathcal{E}^{hs}$ and which is defined in terms of mixed rather than pure states to calculate $S_{cycle}$
\begin{equation}
	S_{cycle}=-\sum_{\alpha}\left( \tilde\lambda_\alpha \textrm{log}_{2}\left( \tilde\lambda_{\alpha}\right)\right),
\end{equation}
where $\tilde\lambda_{\alpha}\equiv\lvert \lambda_{\alpha}\rvert/(\sum_{\alpha}\lvert \lambda_\alpha\rvert)$ are the absolute values of the eigenvalues of $(\sigma\otimes\sigma)\Upsilon$. A cycle entropy $S_{cycle}\approx0$ indicates that there are no (or negligible) internal correlations associated to the bond environment and in this case an optimal or near optimal truncation can be achieved by transforming to WTG and discarding small WTG coefficients. However, if $S_{cycle}$ is larger $(S_{cycle}\gtrapprox 10^{-3})$ (see \cite{evenbly2018gauge}), such a straightforward truncation scheme will not give an optimal truncation and internal correlations may accumulate as the algorithm progresses. We find that in most cases, when starting with a product state, $S_{cycle}$ quickly increases and the FET scheme is required.

\section*{Acknowledgements}
M.H.S. gratefully acknowledges financial support from EPSRC (Grants no. EP/R04399X/1 and no. EP/K003623/2). This work was supported by the Engineering and Physical Sciences Research Council (Grant no. EP/L015242/1).

\bibliography{ipepo_sub.bib}

\end{document}